\documentclass[journal=jcisd8,manuscript=note]{achemso}
\setkeys{acs}{articletitle = true}
\SectionNumbersOn
\usepackage[version=3]{mhchem} 
\usepackage{amsfonts}
\usepackage{makecell}
\usepackage{graphicx}
\usepackage{caption}
\usepackage{hyperref}
\usepackage{placeins}
\usepackage{longtable}
\usepackage{url}
\usepackage{enumerate}
\usepackage[export]{adjustbox}
\usepackage{float}
\usepackage{graphicx}
\usepackage{subfigure}
\usepackage{framed}
\usepackage{amsthm}
\usepackage{appendix}
\usepackage{amssymb}
\usepackage{amsfonts}
\usepackage{enumerate}
\usepackage{algpseudocode}
\usepackage{dsfont}
\usepackage[utf8]{inputenc}
\usepackage{amsmath,amsthm,amssymb}
\usepackage{graphicx}
\usepackage{indentfirst}
\usepackage{rotating}
\usepackage{caption}

\usepackage{seqsplit}

\newcommand{\tabincell}[2]{\begin{tabular}{@{}#1@{}}#2\end{tabular}}

\author{Wenhao Gao}
\affiliation{Department of Chemical and Biomolecular Engineering, Johns Hopkins University, Baltimore, MD}

\author{Sai Pooja Mahajan}
\affiliation{Department of Chemical and Biomolecular Engineering, Johns Hopkins University, Baltimore, MD}

\author{Jeremias Sulam}
\affiliation{Department of Biomedical Engineering, Johns Hopkins University, Baltimore, MD}

\author{Jeffrey J. Gray}
\email{jgray@jhu.edu}
\affiliation{Department of Chemical and Biomolecular Engineering, Johns Hopkins University, Baltimore, MD}

\title{Deep Learning in \\ Protein Structural Modeling and Design}


\begin{document}
\begin{abstract}

Deep learning is catalyzing a scientific revolution fueled by big data, accessible toolkits, and powerful computational resources, impacting many fields including protein structural modeling. Protein structural modeling, such as predicting structure from amino acid sequence and evolutionary information, designing proteins toward desirable functionality, or predicting properties or behavior of a protein, is critical to understand and engineer biological systems at the molecular level. In this review, we summarize the recent advances in applying deep learning techniques to tackle problems in protein structural modeling and design. We dissect the emerging approaches using deep learning techniques for protein structural modeling, and discuss advances and challenges that must be addressed. We argue for the central importance of structure, following the ``sequence $\rightarrow$ structure $\rightarrow$ function'' paradigm. This review is directed to help both computational biologists to gain familiarity with the deep learning methods applied in protein modeling, and computer scientists to gain perspective on the biologically meaningful problems that may benefit from deep learning techniques.

    





\end{abstract}

\tableofcontents


\section{Introduction}

Proteins are linear polymers that fold into various specific conformations to function. The incredible variety of three-dimensional structures determined by the combination and order in which twenty amino acids thread the protein polymer chain (sequence of the protein) enables the sophisticated functionality of proteins responsible for most biological activities. Hence, obtaining the structures of proteins is of paramount importance in both understanding the fundamental biology of health and disease and developing therapeutic molecules. While protein structure is primarily determined by sophisticated experimental techniques such as X-ray crystallography\cite{Slabinski2007}, NMR spectroscopy\cite{Markwick2008} and, increasingly, cryo-electron microscopy\cite{Jonic2009}, computational structure prediction from the genetically encoded amino acid sequence of a protein has been employed as an alternative when  experimental approaches are limited. Computational methods have been used to predict the structure of proteins,\cite{kryshtafovych2019critical} illustrate the mechanism of biological processes,\cite{hollingsworth2018molecular} and determine the properties of proteins\cite{Ranjan2019}. Further, all naturally occurring proteins are a result of an evolutionary process of random variants arising under various selective pressures. Through this process, nature has explored only a small subset of theoretically possible protein sequence space. To explore a broader sequence and structural space that potentially contains proteins with enhanced or novel properties, techniques such as \emph{de novo} design can be employed to generate new biological molecules that have the potential to tackle many outstanding challenges in biomedicine and biotechnology.\cite{Huang2016, yang2019machine} 

\begin{figure}[!t]
    \centering
    \includegraphics[width=\textwidth]{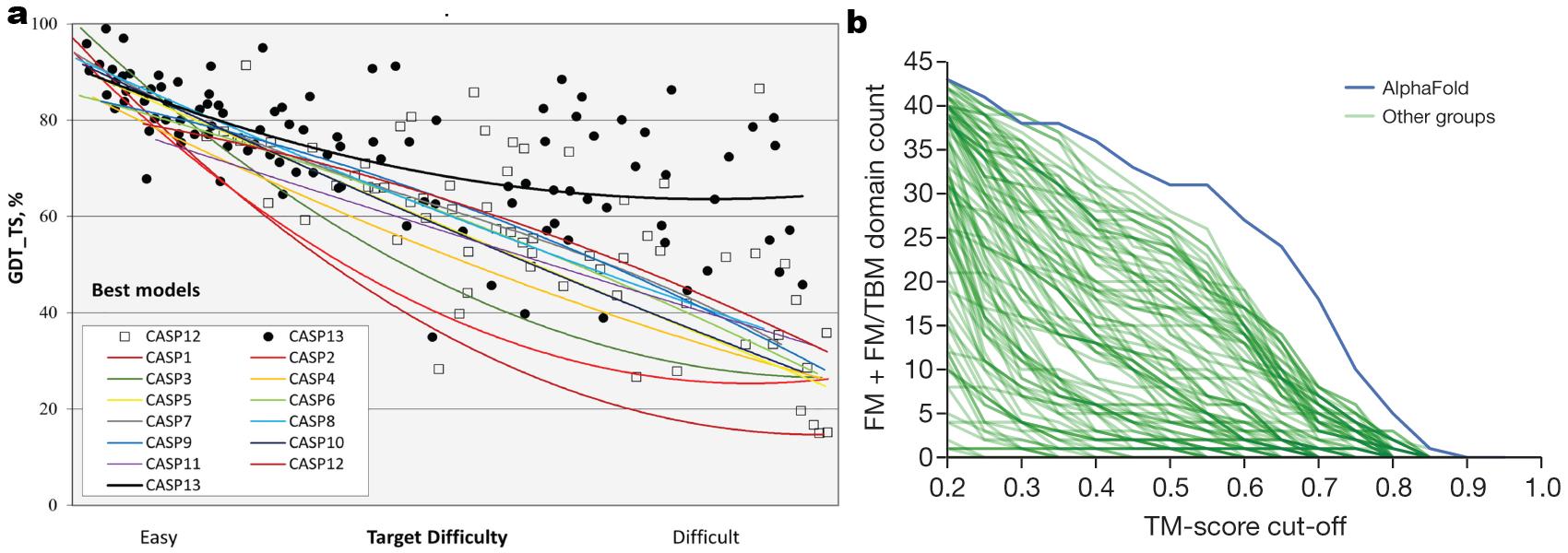}
    \caption{(a) Trend lines of backbone accuracy for the best models in each of the 13 CASP experiments. Individual target points are shown for the two most recent experiments. The accuracy metric, GDT\_TS, is a multiscale indicator of the closeness of the C$\alpha$ atoms in a model to those in the corresponding experimental structure (higher numbers are more accurate). Target difficulty is based on sequence and structure similarity to other proteins with known experimental structures (see \citeauthor{kryshtafovych2019critical} \cite{kryshtafovych2019critical} for details). 
    Figure from \citeauthor{kryshtafovych2019critical}\ (2019)\cite{kryshtafovych2019critical}. (b) Number of FM+FM/TBM (FM: free modeling, TBM: template-based modeling) domains (out of 43) solved to a TM-score threshold for all groups in CASP13. AlphaFold ranked 1st among them, showing the progress is mainly due to the development of DL based methods. 
    Figure from \citeauthor{senior2020improved} (2020)\cite{senior2020improved}}
    \label{fig:casp13}
\end{figure}

While the application of machine learning and more general statistical methods in protein modeling can be traced back decades,\cite{bohr1990novel, Schneider1994, Schneider1998, Ofran2003, Nielsen2003} recent advances in machine learning, especially in deep learning (DL)\cite{lecun2015deep} related techniques, have opened up new avenues in many areas of protein modeling. \cite{review-Angermueller2016, ching2018opportunities, mura2018structural, Noe2019} DL is a set of machine learning techniques based on stacked neural network layers that  parameterize functions in terms of compositions of affine transformations and non-linear activation functions. Their ability to extract domain-specific features that are adaptively learned from data for a particular task often enables them to surpass the performance of more traditional methods. DL has made dramatic impacts on digital applications like image classification\cite{guo2016deep}, speech recognition\cite{young2018recent} and game playing\cite{silver2017mastering}. Success in these areas has inspired an increasing interest in more complex data types, including protein structures.\cite{Senior2019} In the most recent Critical Assessment of Structure Prediction (CASP13 held in 2018),\cite{kryshtafovych2019critical} a biennial community experiment to determine the state-of-the-art in protein structure prediction, DL-based methods accomplished a striking improvement in model accuracy  (see Figure \ref{fig:casp13}), especially in the ``difficult'' target category where comparative modeling (starting with a known, related structure) is ineffective. The CASP13 results show that the complex mapping from amino acid sequence to three-dimensional protein structure can be successfully learned by a neural network and generalized to unseen cases. Concurrently, for the protein design problem, progress in the field of deep generative models has spawned a range of promising approaches.\cite{ingraham2019generative, anand2018generative, OConnell2018} 

\begin{figure}[h!]
    \centering
    \includegraphics[width=\textwidth]{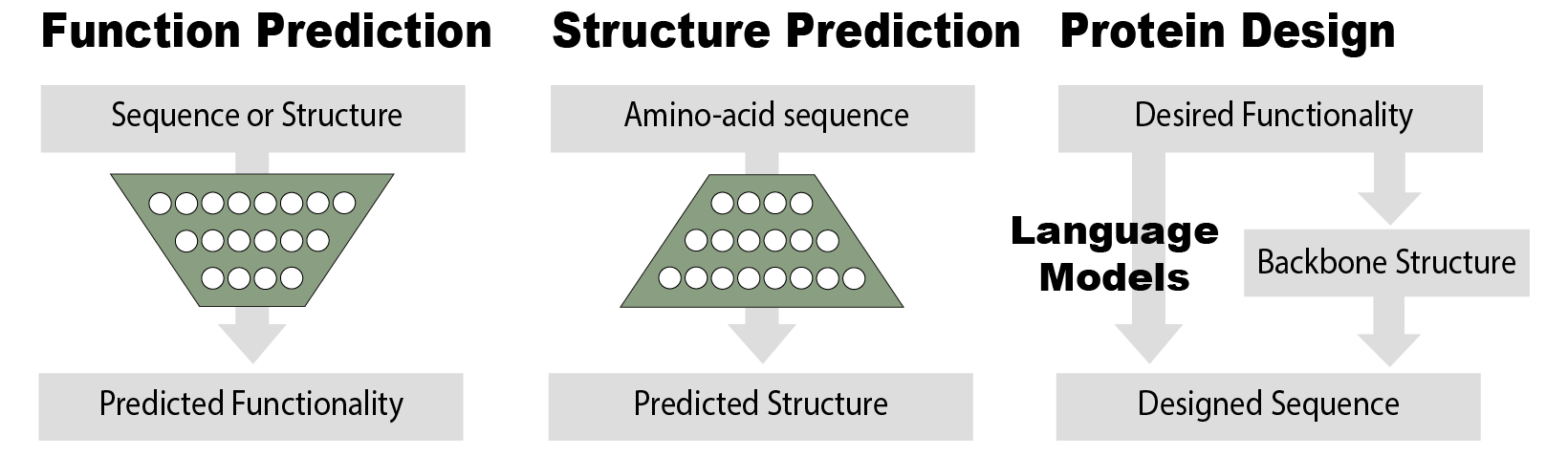}
    \caption{Schematic comparison of three major tasks in protein modeling: function prediction, structure prediction and protein design. In function prediction, the sequence and/or the structure is known and the functionality is needed as output of a neural net. In structure prediction, sequence is known input and structure is unknown output. Protein design starts from desired functionality, or a step further, structure that can perform this functionality. The desired output is a sequence that can fold into the structure or has such functionality.}
    \label{fig:schematic}
\end{figure}

In this review, we summarize the recent progress in applying DL techniques to the problem of protein modeling and discuss the potential pros and cons. We limit our scope to protein structure and function prediction, protein design with DL (see Figure \ref{fig:schematic}), and a wide array of popular frameworks used in these applications. We discuss the importance of protein representation, and summarize the approaches to protein design based on DL for the first time. We also emphasize the central importance of protein structure, following the sequence $\rightarrow$ structure $\rightarrow$ function paradigm and argue that approaches based on structures may be most fruitful. We refer the reader to other review papers for more information on applications of DL in biology and medicine\cite{ching2018opportunities, review-Angermueller2016}, bioinformatics\cite{li2019deep}, structural biology\cite{mura2018structural}, folding and dynamics\cite{Noe2019, noe2020machine}, antibody modeling,\cite{graves2020review} and structural annotation and prediction of proteins.\cite{kandathil2019recent, le2020deep} 


\section{Protein structure prediction and design}

\subsection{Problem definition}

The prediction of protein three-dimensional structure from amino acid sequence has been a grand challenge in computational biophysics for decades.\cite{pauling1939structure, Kuhlman2019} Folding of peptide chains is a fundamental concept in biophysics, and atomic-level structures of proteins and complexes are often the starting point to understand their function and to modulate or engineer them. Thanks to the recent advances in next-generation sequencing technology, there are now over 180 million protein sequences recorded in UniProt dataset.\cite{uniprot2019uniprot} In contrast, only 158,000 experimentally determined structures are available in the Protein Data Bank (PDB). Thus, computational structure prediction is a critical problem of both practical and theoretical interest.

More recently, the advances in structure prediction have led to an increasing interest in the protein design problem. In design, the objective is to obtain a novel protein sequence that will fold into a desired structure or perform a specific function, such as catalysis. Naturally occurring proteins represent only an infinitesimal subset of all possible amino acid sequences selected by the evolutionary process to perform a specific biological function.\cite{Huang2016}  Proteins with more robustness (higher thermal stability, resistance to degradation) or enhanced properties (faster catalysis, tighter binding) might lie in the space that has not been explored by nature, but is potentially accessible by \emph{de novo} design. The current approach for computational \emph{de novo} design is based on physical and evolutionary principles and requires significant domain expertise. Some successful examples include novel folds, \cite{kuhlman2003design} enzymes,\cite{donnelly2018novo} vaccines,\cite{correia2014proof} novel protein assemblies,\cite{king2012computational} ligand-binding protein, \cite{tinberg2013computational} and membrane proteins.\cite{joh2014novo}

\subsection{Conventional computational approaches}

The current methodology for computational protein structure prediction is largely based on Anfinsen's thermodynamic hypothesis\cite{Anfinsen1973}, which states that the native structure of a protein must be the one with the lowest free-energy, governed by the energy landscape of all possible conformations associated with its sequence. Finding the lowest-energy state is challenging because of the immense space of possible conformations available to a protein, also known as the ``sampling problem'' or Levinthal's paradox\cite{Levinthal1968}. Furthermore, the approach requires accurate free energy functions to describe the protein energy landscape and rank different conformations based on their energy, referred as the ``scoring problem''. In light of these challenges, current computational techniques rely heavily on multi-scale approaches. Low-resolution, coarse-grained energy functions are employed to capture large scale conformational sampling such as the hydrophobic burial and formation of local secondary structural elements. Higher-resolution energy functions are employed to explicitly model finer details such as amino acid side-chain packing, hydrogen bonding and salt bridges.\cite{Li2018finding}  

Protein design problems, sometimes known as the inverse of structure prediction problems, require a similar toolbox. Instead of sampling the conformational space, a protein design protocol samples the sequence space that folds into the desired topology. Current efforts can be broadly divided into two classes: modifying an existing protein with known sequence and properties, or generating novel proteins with sequences unrelated to those found in nature. The former protocol evolves an existing protein's amino acid sequence (and as a result, structure and properties), and the latter is called \emph{de novo} protein design. 



Despite significant progress in the last several decades in the field of computational protein structure prediction and design\cite{Huang2016,Kuhlman2019}, accurate structure prediction and reliable design both remain challenging. Conventional approaches rely heavily on the accuracy of the energy functions to describe protein physics and the efficiency of sampling algorithms to explore the immense protein sequence and structure space. 

\section{Deep learning architectures}

In conventional computational approaches, predictions from data are made by means of physical equations and modeling. Machine learning puts forward a different paradigm in which algorithms automatically infer -- or \emph{learn} -- a relationship between inputs and outputs from a set of hypotheses. Consider a collection of $N$ training samples comprising features $\boldsymbol{x}$ in an input space $\mathcal{X}$ (\textit{e.g.}, amino acid sequences), and corresponding labels $y$ in some output space $\mathcal{Y}$ (\textit{e.g.}, residue pairwise distances), where $\{\boldsymbol{x_i},y_i\}_{i=1}^N$ are sampled independently and identically distributed from some joint distribution $\mathcal P$. Additionally, consider a function $f:\mathcal{X}\to\mathcal{Y}$ in some function class $\mathcal{H}$, and a loss function $\ell:\mathcal{Y}\times\mathcal{Y} \to \mathbb R$ that measures how much $f(\boldsymbol{x})$ deviates from the corresponding label $y$. The goal of supervised learning is to find a function $f\in\mathcal H$ which minimizes the expected loss, $\mathbb E[\ell(f(\boldsymbol{x}),y)]$, for $(\boldsymbol{x},y)$ sampled from $\mathcal P$. Since one does not have access to the true distribution but rather $N$ samples from it, the popular Empirical Risk Minimization (ERM) approach seeks to minimize the loss over the training samples instead. 
In neural network models, in particular, the function class is parameterized by a collection of weights. Denoting these parameters collectively by $\boldsymbol{\theta}$, ERM boils down to an optimization problem of the form
\begin{equation} \label{eq:opt_problem}
    \min_{\boldsymbol{\theta}} \frac{1}{N}\sum_{i=1}^N \ell(f_{\boldsymbol{\theta}} (\boldsymbol{x_i}),y_i).
\end{equation}

The choice of the network determines how the hypothesis class is parameterized. Deep neural networks typically implement a non-linear function as the composition of affine maps, $\mathcal{W}_l: \mathbb{R}^{n_l} \to \mathbb{R}^{n_{l+1}}$, where  $\mathcal{W}_l \boldsymbol{x} = W_l \boldsymbol{x} + \boldsymbol{b_l}$, and other non-linear activation functions, $\sigma(\cdot)$. REctifying Linear Units (RELU) and max-pooling are some of the most popular non-linear transformations applied in practice. The architecture of the model determines how these functions are composed, the most popular option being their sequential composition $f(\boldsymbol{x}) = \mathcal{W}_L~\sigma(\mathcal{W}_{L-1}~\sigma(\mathcal{W}_{L-2}~\sigma(\dots \mathcal{W}_2 \sigma(\mathcal{W}_1 \boldsymbol{x}) )))$ 
for a network with $L$ layers. Computing $f(\boldsymbol{x})$ is typically referred to as the \emph{forward pass}. 

\begin{figure}[h!]
    \centering
    \includegraphics[width=\textwidth]{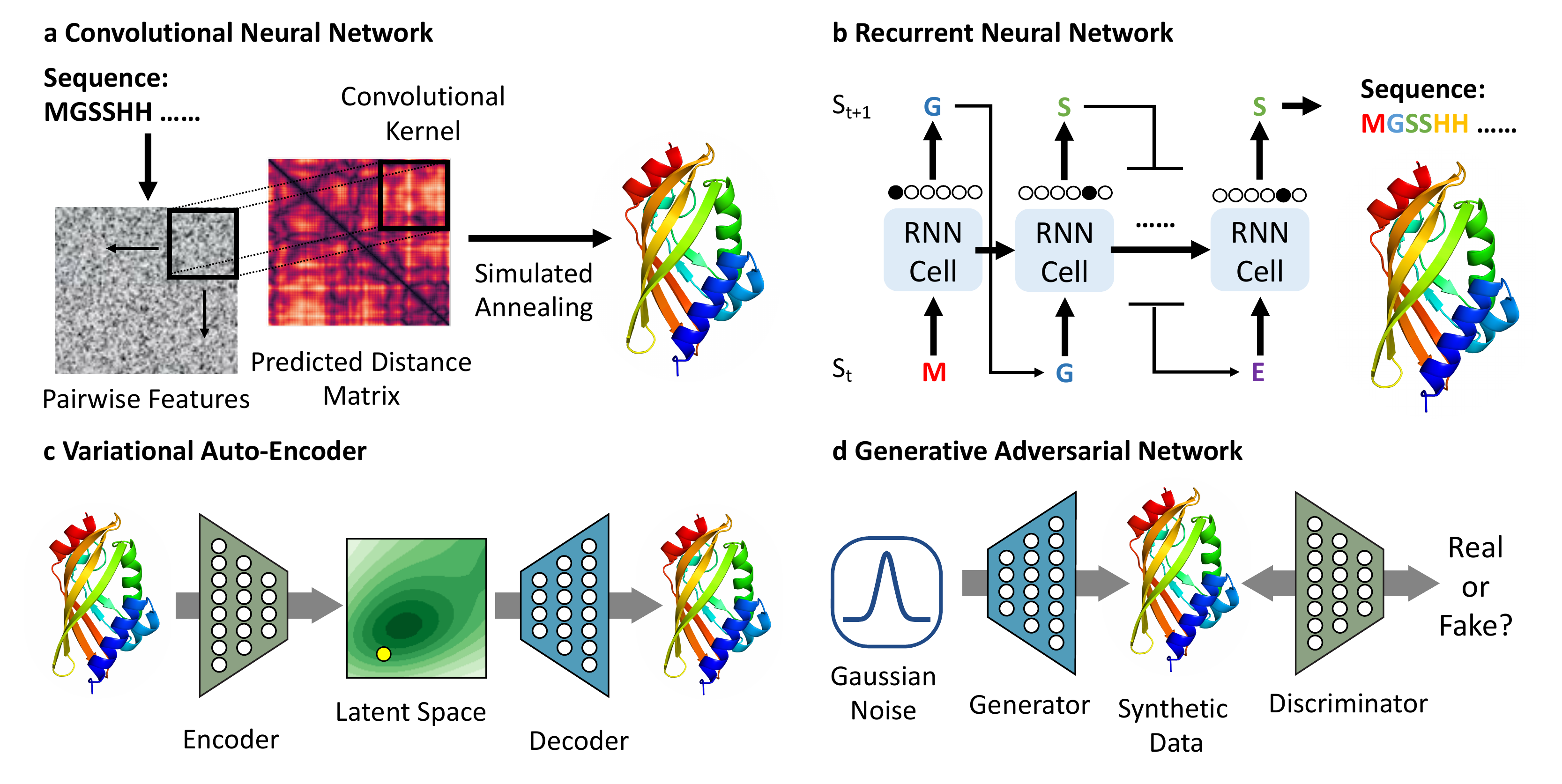}
    \caption{Schematic representation of several architectures used in protein modeling and design. (a) CNNs are widely used in structure prediction. (b) RNNs learn in an auto-regressive way and can be used for sequence generation. (c) The VAE can be jointly trained by protein and properties to construct a latent space correlated with properties. (d) In the GAN setting, a mapping from \emph{a priori} distribution to the design space can be obtained via the adversarial training.}
    \label{fig:models}
\end{figure}

We will not dwell on the details of the optimization problem in Eq. \eqref{eq:opt_problem}, which is typically carried out via stochastic gradient descent algorithms or variations thereof, efficiently implemented via \emph{back-propagation}. Rather, in this section we summarize some of the most popular models widely used in protein structural modeling. High-level diagrams of the major architectures are shown in Figure \ref{fig:models}.

\subsection{Convolutional neural networks (CNN)}

Convolutional networks architectures \cite{lecun1990handwritten} are most commonly applied to image analysis or other problems where shift-invariance or co-variance is needed. Inspired by the fact that an object on an image can be shifted in the image and still be the same object, CNNs adopt convolutional kernels for the layer-wise affine transformation to capture this translational invariance. A 2D convolutional kernel $\mathbf{w}$ applied to a 2D image data $\boldsymbol{x}$ can be defined as:
\begin{equation}
   \boldsymbol{S}(i, j) = (\boldsymbol{x}*\mathbf{w})(i, j) = \sum_{m} \sum_{n} \boldsymbol{x}(m, n)\mathbf{w}(i - m, j - n)
\end{equation}
where $\boldsymbol{S}(i, j)$ represents the output at position $(i, j)$, $\boldsymbol{x}(m, n)$ is the value of the input $\boldsymbol{x}$ at position $(m, n)$, $\mathbf{w}(i - m, j - n)$ is the parameter of kernel $\mathbf{w}$ at position $(i-m, j-n)$, and the summation is over all possible positions. An important variant of CNN is the residual network (ResNet) \cite{he2016deep}, which incorporates skip-connections between layers. These modification have shown great advantages in practice, aiding the optimization of these typically huge models. CNNs, especially ResNets, have been widely used in protein structure prediction. An example is AlphaFold \cite{Senior2019}, in which the input is given by residue pair-wise features and the output is a corresponding residue distance map (Figure \ref{fig:models}a).

\subsection{Recurrent neural networks (RNN)}

Recurrent architectures are based on applying several iterations of the same function along a sequential input \cite{jordan1986serial}. This can be seen as an \emph{unfolded} architecture, and has been widely used to process sequential data, such as written text and time series data. An example of an RNN approach in the context of protein prediction is using an N-terminal subsequence of a protein and predicting the next amino acid in the protein (Figure \ref{fig:models}b; \textit{e.g.} \citeauthor{Muller2018}\cite{Muller2018}). With an initial hidden state $\boldsymbol{h}^{(0)}$ and sequential data [$\boldsymbol{x}^{(1)}, \boldsymbol{x}^{(2)}, \dots, \boldsymbol{x}^{(n)}$], we can obtain hidden states recursively:
\begin{equation}
    \boldsymbol{h}^{(t)} = g^{(t)} (\boldsymbol{x}^{(t)}, \boldsymbol{x}^{(t-1)}, \boldsymbol{x}^{(t-2)}, \dots, \boldsymbol{x}^{(1)}) = f(\boldsymbol{h}^{(t-1)}, \boldsymbol{x}^{(t)}; \boldsymbol{\theta}),
\end{equation}
where $f$ represents a function or transformation from one position to the next, and $g^{(t)}$ represents the accumulative transformation up to position $t$. The hidden state vector at position $i$, $\boldsymbol{h}^{(i)}$, contains all the information that has been seen before. As the same set of parameters (usually called a cell) can be applied recurrently along the sequential data, an input of variable length can be fed to an RNN. Due to the gradient vanishing and explosion problem (the error signal decreases or increases exponentially during training), more recent variants of standard RNN, namely Long Short-Term Memory (LSTM) \cite{hochreiter1997long} and Gated Recurrent Unit (GRU) \cite{cho2014learning} are more widely used. 

\subsection{Variational auto-encoder (VAE)}

Auto-Encoders \cite{hinton1994autoencoders}(AEs), unlike the ones discussed so far, provide a model for \emph{unsupervised} learning. Within this unsupervised framework, an auto-encoder does not learn labeled outputs but instead attempts to learn some representation of the original input. 
This is typically accomplished by training two parametric maps: an \emph{encoder} function $g:\mathcal{X}\to\mathbb{R}^m$ that maps an input $\boldsymbol{x}$ to an $m$-dimensional representation or latent space, and a \emph{decoder} $g:\mathbb{R}^m\to\mathcal{X}$ intended to implement the inverse map so that $f(g(\boldsymbol{x}))\approx \boldsymbol{x}$. Typically, the latent representation is of small dimension ($m$ is smaller than the ambient dimension of $\mathcal X$) or constrained in some other way (\textit{e.g.},\ through sparsity). 

Variational Auto-Encoders (VAEs) \cite{kingma2013auto, kingma2019introduction}, in particular,
provide a stochastic map between the input space and the latent space. This is beneficial because, while the input space may have a highly complex distribution, the distribution of the representation $\boldsymbol{z}$ can be much simpler; \textit{e.g.},\ Gaussian. These methods are derived from variational inference, a method from machine learning that approximates probability densities through optimization \cite{blei2017variational}. 
The stochastic encoder, given by the \emph{inference model} $q_{\boldsymbol{\phi}} (\boldsymbol{z} | \boldsymbol{x})$ and parametrized by weights $\boldsymbol{\phi}$, is trained to approximate the true posterior distribution of the representation given the data, $p_{\boldsymbol{\theta}}(\boldsymbol{z}|\boldsymbol{x})$. The decoder, on the other hand, provides an estimate for the data given the representation, $p_{\boldsymbol{\theta}} (\boldsymbol{x} | \boldsymbol{z})$. Direct optimization of the resulting objective is intractable, however. Thus, training is done by maximizing the ``Evidence Lower BOund'' (ELBO), $\mathcal{L}_{\boldsymbol{\theta}, \boldsymbol{\phi}}(\boldsymbol{x})$, instead, which provides a lower bound on the log-likehood of the data:
\begin{equation}
    \mathcal{L}_{\boldsymbol{\theta}, \boldsymbol{\phi}}(\boldsymbol{x}) = \mathds{E}_{\boldsymbol{z} \sim q_{\boldsymbol{\phi}}(\boldsymbol{z}|\boldsymbol{x})}  \log p_{\boldsymbol{\theta}} (\boldsymbol{x}|\boldsymbol{z}) - D_{KL}\left(q_{\boldsymbol{\phi}}(\boldsymbol{z}|\boldsymbol{x}) ~||~ p_{\boldsymbol{\theta}}(\boldsymbol{z}|\boldsymbol{x})\right).
\end{equation}
Here, $D_{KL}(q_{\boldsymbol{\phi}}||p_{\boldsymbol{\theta}})$ is the Kullback–Leibler (KL) divergence, which quantifies the distance between distributions $q_{\boldsymbol{\phi}}$ and $p_{\boldsymbol{\theta}}$. Employing Gaussians for the factorized variational and likelihood distributions, as well as employing a change of variables via differentiable maps, allows for the efficient optimization of these architectures. 


An example of applying VAE in the protein modeling field is learning a representation of anti-microbial protein sequences (Figure \ref{fig:models}c; \textit{e.g.}, \citeauthor{Das2018}\cite{Das2018}). The resulting continuous real-valued representation can then be used to generate new sequences likely to have antimicrobial properties.

\subsection{Generative adversarial network (GAN)}

Generative Adversarial Networks (GANs) \cite{goodfellow2014generative} are another class of unsupervised (generative) models. Unlike VAEs, GANs are trained by an adversarial game between two models, or networks: a \emph{generator}, $G$, which given a sample, $\boldsymbol{z}$, from some simple distribution $p_{\boldsymbol{z}}(\boldsymbol{z})$ (\textit{e.g.}, Gaussian), seeks to map it to the distribution of some data class (\textit{e.g.}, naturally looking images); and a \emph{discriminator}, $D$, whose task is to detect whether the images are real (\textit{i.e.}, belonging to the true distribution of the data, $p_{data}(\boldsymbol{x})$), or fake (produced by the generator). With this game-based setup, the generator model is trained by maximizing the error rate of the discriminator, thereby training it to ``fool'' the discriminator. The discriminator, on the other hand, is trained to foil such fooling. The original objective function as formulated in \cite{goodfellow2014generative} is:
\begin{equation}
    \min_{G}\max_{D} V(D, G) = \mathds{E}_{\boldsymbol{x} \sim p_{data} (\boldsymbol{x})} [\log D(\boldsymbol{x})] + \mathds{E}_{\boldsymbol{z} \sim p_{\boldsymbol{z}}(\boldsymbol{z})} [\log (1 - D(G(\boldsymbol{z})))].
\end{equation}
Training is performed by stochastic optimization of this differentiable loss function. While intuitive, this original GAN objective can suffer from issues such as mode collapse and instabilities during training. The Wasserstein GAN (WGAN) \cite{arjovsky2017wasserstein} is a popular extension of GAN which introduces a Wasserstein-1 distance measure between distributions, leading to easier and more robust training in practice.\cite{kurach2018large} An example of GAN, in the context of protein modeling, is the work of \citeauthor{anand2019fully}, to learn the distribution of protein backbone distances and generate novel protein-like folds (Figure \ref{fig:models}d).\cite{anand2019fully} One network $G$ generates folds, and a second network $D$ aims to distinguish between generated folds and fake folds.

\section{Protein representation and function prediction} 

One of the most fundamental challenges in protein modeling is the prediction of functionality from sequence or structure. Function prediction is typically formulated as a supervised learning problem. The property to predict can either be a protein-level property, such as a classification as an enzyme or non-enzyme, \cite{niepert2016learning} or a residue-level property, such as the sites or motifs of phosphorylation (DeepPho) \cite{luo2019deepphos} and polyadenylation (Terminitor). \cite{yang2019termin} The challenging part here and in the following models is how to represent the protein. Representation refers to the encoding of a protein that serves as an input for prediction tasks or the output for generation tasks. Although a deep neural network is in principle capable of extracting complex features, a well-chosen representation can make learning more effective and efficient. \cite{bengio2013representation} In this section, we will introduce the commonly used representations of proteins in DL models: sequence-based, structure-based, and one special form of representation relevant to computational modeling of proteins: coarse-grained models.

\begin{figure}[h!]
    \centering
    \includegraphics[width=\textwidth]{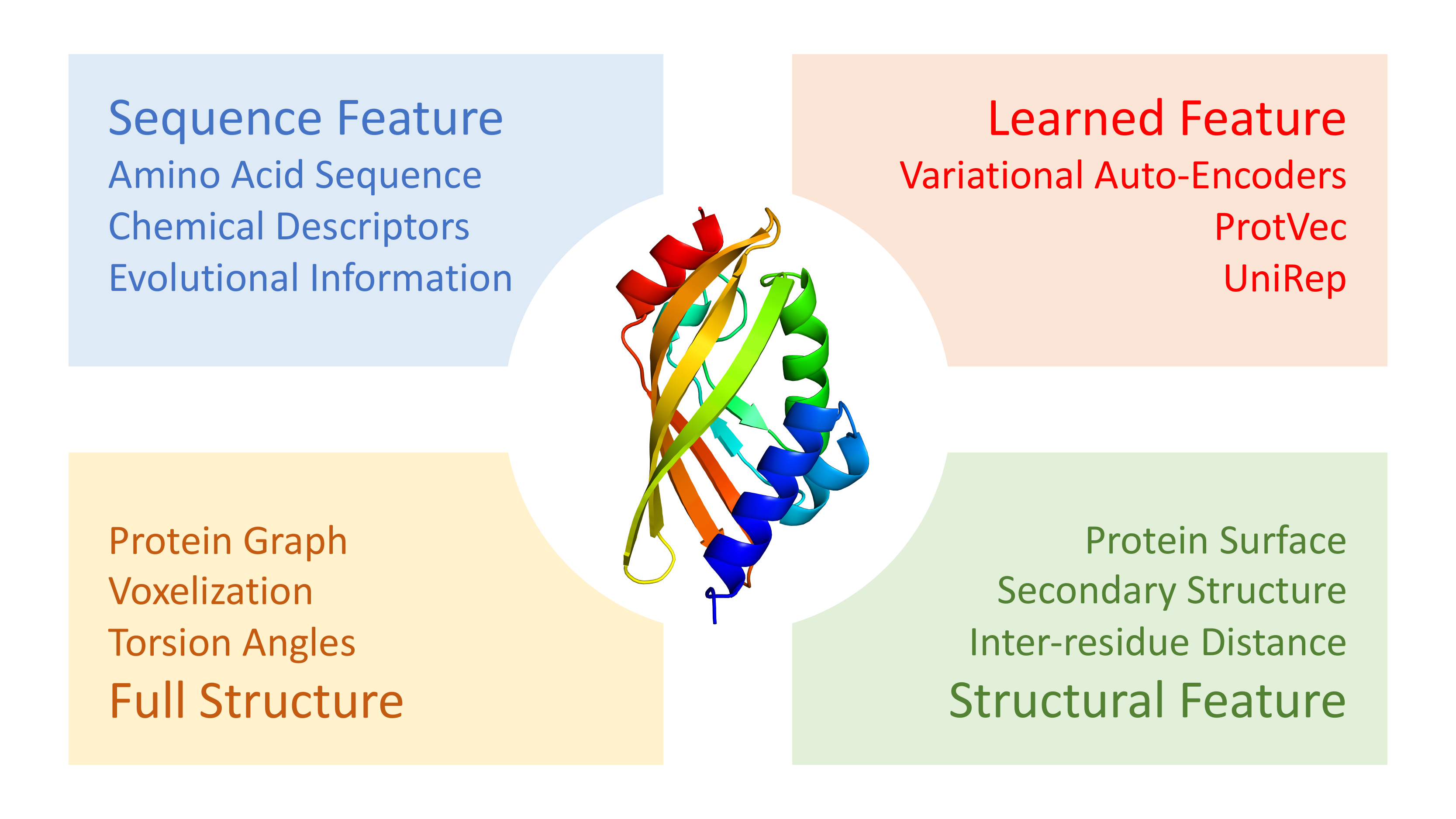}
    \caption{Different types of representation schemes applied to a protein.}
    \label{fig:representation}
\end{figure}

\subsection{Amino acid sequence as representation}

As the amino acid sequence contains the information essential to reach the folded structure for most proteins,\cite{Anfinsen1973} it is widely used as an input in functional prediction and structure prediction tasks. The amino acid sequence, like other sequential data, is typically converted into one-hot encoding based representation (each residue is represented with one high bit to identify the amino acid type and all the others low), that can be directly used in many sequence-based DL techniques. \cite{romero2013navigating, bedbrook2017machine} However, this representation is inherently sparse, and thus, sample-inefficient. There are many easily accessible additional features that can be concatenated with amino acid sequences providing structural, evolutionary, and biophysical information. Some widely used features include predicted secondary structure, high-level biological features such as sub-cellular localization and unique functions\cite{ofer2015profet}, and physical descriptors such as AAIndex,\cite{kawashima2007aaindex} hydrophobicity, ability to form hydrogen bonds, charge, solvent-accessible surface area, etc.
A sequence can be augmented with additional data from sequence databases such as multiple sequence alignments (MSA) or  position-specific scoring matrices (PSSM)\cite{wang2016protein}, or pairwise residue co-evolution features. Table \ref{tab:features} lists typical features as used in CUProtein \cite{drori2019accurate}. 

\subsection{Learned representation from amino acid sequence}

Because the performance of machine learning algorithms highly depends on the features we choose, labor-intensive and domain-based feature engineering was vital for traditional machine learning projects. Now, the exceptional feature extraction ability of neural networks makes it possible to ``learn'' the representation, with or without giving the model any labels. \cite{bengio2013representation} As publicly available sequence data is abundant (See Table \ref{tab:data}), a well-learned representation that utilizes these data to capture more information is of particular interest. The class of algorithms that address the label-less learning problem fall under the umbrella of unsupervised or semi-supervised learning, which extracts information from unlabeled data to reduce the number of labeled samples needed. 

The most straightforward way to learn from amino acid sequence is to directly apply natural language processing (NLP) algorithms. Word2Vec \cite{mikolov2013efficient} and Doc2Vec \cite{le2014distributed} are groups of algorithms widely used for learning word or paragraph embeddings. These models are trained by either predicting a word from its context or predicting its context from one central word. To apply these algorithm, \citeauthor{asgari2015continuous}  first proposed a Word2Vec based model called BioVec that interprets the non-overlapping 3-mer sequence of amino acids (eg. alanine-glutamine-lysine or AQL) as ``words'' and lists of shifted ``words'' as ``sentences''.\cite{asgari2015continuous} They then represent a protein as the summation of all overlapping sequence fragments of length \emph{k}, or $k$-mers (called ProtVec). Predictions based on the ProtVec representation outperformed state-of-the-art machine learning methods in the Pfam protein family \cite{el2019pfam} classification (93\% accuracy for $\sim$ 7,000 proteins, versus 69.1-99.6\% \cite{cai2003svm} and 75\% \cite{aragues2007characterization} for previous methods). Many Doc2Vec-type extensions were developed based on the ``3-mer'' protocol.  \citeauthor{kimothi2016distributed} showed that non-overlapping $k$-mers perform better than the overlapping ones,\cite{kimothi2016distributed} and \citeauthor{yang2018learned}  compared the performance of all Doc2Vec frameworks for thermostability and enantioselectivity prediction.\cite{yang2018learned}

In these approaches, the three-residue segmentation of a protein sequence is arbitrary and does not embody any biophysical meaning. Alternatively, \citeauthor{alley2019unified}  directly used an RNN (unidirectional multiplicative long-short-term-memory or mLSTM\cite{krause2016multiplicative})  model, called UniRep, to summarize arbitrary length protein sequences into a fixed-length real representation by averaging over the representation of each residue.\cite{alley2019unified} Their representation achieved lower mean squared errors on 15 property prediction tasks (absorbance, activity, stability, etc.) compared to former models, including \citeauthor{yang2018learned}'s Doc2Vec. \cite{yang2018learned} Notably, their sequence representation based model, UniRep Fusion, was able to outperform stability ranking predictions made by Rosetta, which uses sequence, structure, and a scoring function trained on various biophysical data. 

Auto-encoders can also provide representations for subsequent supervised tasks.\cite{kingma2013auto} \citeauthor{ding2019deciphering}  showed that a VAE model is able to capture evolutionary relationships between sequences and stability of proteins,\cite{ding2019deciphering} while \citeauthor{Sinai2017}  and \citeauthor{Riesselman2018}  showed that the latent vectors learned from VAEs are able to predict the effects of mutations on fitness and activity for a range of proteins such as poly(A)-binding protein, DNA methyltransferase and $\beta$-lactamase.\cite{Sinai2017,Riesselman2018} Recently,  a lower-dimensional embedding of the sequence was learned for the more complex task of structure prediction. \cite{drori2019accurate} \citeauthor{alley2019unified}'s UniRep surpassed former models, but since UniRep is trained on 24 million sequences and previous models (\textit{e.g.}, Prot2Vec) were trained on much smaller datasets (0.5 million), it is not clear if the improvement was due to better methods or the larger training dataset.\cite{alley2019unified} \citeauthor{rao2019evaluating} introduced multiple biological-relevant semi-supervised learning tasks, TAPE, and benchmarked the performance against various protein representations. Their results show conventional alignment-based inputs still outperform current self-supervised models on multiple tasks, and the performance on a single task cannot evaluate the capacity of models. \cite{rao2019evaluating} A comprehensive and persuasive comparison of representations is required.


\subsection{Structure as representation}
\label{chap:representation_structure}

Since the most important functions of a protein (binding, signaling, catalysis etc.) can be traced back to the 3D structure of the protein, direct use of 3D structural information, and analogously, learning a good representation based on 3D structure, are highly desired. The direct use of raw 3D representations (such as coordinates of atoms) is hindered by considerable challenges, including the processing of unnecessary information due to translation, rotation, and permutation of atomic indexing. 
\citeauthor{Townshend2018}\cite{Townshend2018} and \citeauthor{Simonovsky2019}\cite{Simonovsky2019}, obtained a translationally-invariant, 3D representation of each residue by voxelizing its atomic neighborhood for a grid-based 3D CNN model. Alternatively, the torsion angles of the protein backbone, which are invariant to translation and rotation, can fully recapitulate protein backbone structure under the common assumption that variation in bond lengths and angles is negligible. \citeauthor{AlQuraishi2019}  employed backbone torsion angles to represent the 3D structure of the protein as a 1D data vector.\cite{AlQuraishi2019} However, because a change in a backbone torsion angle at a residue affects the inter-residue distances between all preceding and subsequent residues, these 1D variables are highly interdependent, which can frustrate learning. To circumvent these limitations, many approaches use 2D projections of 3D protein structure data such as residue-residue distance and contact maps \cite{anand2018generative, wang2017accurate} and pseudo-torsion angles and bond angles that capture the relative orientations between pairs of residues \cite{yang2019improved}. While these representations guarantee translational and rotational invariance, they do not guarantee invertibility back to the 3D structure. The structure must be reconstructed by applying constraints on distance or contact parameters using algorithms such as gradient descent minimization, multidimensional scaling, a program like the Crystallography and NMR system (CNS), \cite{brunger2007version} or in conjunction with an energy-function-based protein structure prediction program. \cite{Senior2019}

An alternative to the above approaches for representing protein structures is the use of a graph, \textit{i.e.}, a collection of nodes or vertices connected by edges. Such a representation is highly amenable to the Graph Neural Network (GNN) paradigm, \cite{zhou2018graph} which has recently emerged as a powerful framework for non-Euclidean data\cite{Ahmed2018DeepLA} in which the data are represented with relationships and inter-dependencies, or edges, between objects, or nodes \cite{wu2019comprehensive}. While the representation of proteins as graphs and the application of graph theory to study their structure and properties has a long history, \cite{vishveshwara2002protein} the efforts to apply GNNs to protein modeling and design is quite recent. As a benchmark, many GNNs \cite{niepert2016learning, ying2018hierarchical} have been applied to classify enzymes from non-enzymes in the PROTEINS\cite{borgwardt2005protein} and D\&D \cite{dobson2003distinguishing} datasets. \citeauthor{fout2017protein}  utilized a GNN in developing a model for protein-protein interface prediction.\cite{fout2017protein} In their model, the node feature comprised residue composition and conservation, accessible surface area, residue depth, and protrusion index; and the edge feature comprised a distance and an angle between the normal vectors of the amide plane of each node/residue. A similar framework was used to predict antibody-antigen binding interfaces \cite{bailey2019learning}. \citeauthor{zamora2019structural} \cite{zamora2019structural} and \citeauthor{gligorijevic2019structure} \cite{gligorijevic2019structure} further generalized and validated the use of graph-based representations and the Graph Convolutional Network (GCN) framework in protein function prediction tasks, using a Class Activation Map (CAM) to interpret the structural determinants of the functionalities. \citeauthor{torng2019graph}  applied GCNs to model pocket-like cavities in proteins to predict the interaction of proteins  with small molecules,\cite{torng2019graph}  and \citeauthor{ingraham2019generative}  adopted a graph based transformer model to perform a protein sequence design task.\cite{ingraham2019generative} These examples demonstrate the generality and potential of the graph-based representation and GNNs to encode structural information for protein modeling.

The surface of the protein or a cavity is an information-rich region that encodes how a protein may interact with other molecules and its environment.
Recently, \citeauthor{gainza2019deciphering} \cite{gainza2019deciphering} used a geometric DL framework \cite{bronstein2017geometric} to learn a surface-based representation of the protein, called MaSIF. They calculated ``fingerprints'' for patches on the protein surface using geodesic convolutional layers, which were further used to perform tasks such as binding site prediction or ultra-fast protein-protein interaction (PPI) search. The performance of MaSIF approached  the baseline of current methods in docking and function prediction, providing a proof-of-concept to inspire more applications of geometry-based representation learning.

\subsection{Score function and force field}

A high-quality force field (or, more generally, score function) for sampling and/or ranking models (decoys) is one of the most vital requirements for protein structural modeling. \cite{nerenberg2018new} A force field describes the potential energy surface of a protein. A score function may contain knowledge-based terms that do not necessarily have a valid physical meaning, and they are designed to distinguish near-native conformations from non-native ones (for example, learning the GDT\_TS \cite{derevyanko2018deep}). A molecular dynamics (MD) or Monte Carlo (MC) simulation with a state-of-the-art force field or score function can reproduce reasonable statistical behaviors of biomolecules. \cite{best2012optimization, weiner1984new, alford2017rosetta}

Current DL-based efforts to learn the force field can be divided into two classes: ``fingerprint''-based and graph-based. \citeauthor{behler2007generalized} developed roto-translationally invariant features, \textit{i.e.}, Behler–Parrinello fingerprint, to encode the atomic environment for neural networks to learn potential surfaces from Density Functional Theory (DFT) calculations.\cite{behler2007generalized} \citeauthor{smith2017ani} extended this framework and tested its accuracy by simulating systems up to 312 atoms (Trp-cage) for 1 ns. \cite{smith2017ani, smith2018less} Another family that includes deep tensor neural network (DTNN) \cite{schutt2017quantum} and SchNet, \cite{schutt2018schnet} utilizes graph convolution to learn a representation for each atom within its chemical environment. 
Though the prediction quality and the ability to learn a representation with novel chemical insight make the graph-based approach increasingly popular, \cite{noe2020machine} the application has mainly focused on small organic molecules as it scales poorly to larger systems. 

A shift towards DL-based score functions is anticipated, especially due to the enormous gains in speed and efficiency. For example, \citeauthor{zhang2018deep} showed that MD simulation on a neural potential was able to reproduce energies, forces, and time-averaged properties comparable to \textit{ab initio} MD (AIMD) at a cost that scales linearly with system size, compared to cubic scaling typical for AIMD with DFT. \cite{zhang2018deep}  Though these force fields are, in principle, generalizable to larger systems, direct applications of neural potential to model full proteins are still rare. PhysNet, trained on a set of small peptide fragments (at most eight heavy atoms), was able to generalize to deca-alanine (Ala10),\cite{unke2019physnet} and ANI-1x and AIMNet have been tested on Chignolin (10 residues) and Trp-cage (20 residues) within the ANI-MD benchmark dataset. \cite{smith2018less, zubatyuk2019accurate} \citeauthor{lahey2020simulating} and \citeauthor{wang2020combining} combined the Quantum Mechanics/Molecular Mechanics (QM/MM) strategy \cite{senn2009qm} and the neural potential to model docking with small ligands and larger proteins (up to 82 residues). \cite{lahey2020simulating, wang2020combining}


\subsection{Coarse-grained models}

Coarse-grained models are higher-level abstractions of biomolecules, such as using a single pseudo-atom or a bead to represent multiple atoms, grouped based on local connectivity and/or chemical properties. Coarse-graining smoothens out the energy landscape, and thereby helps avoid trapping in local minima and speeds up conformational sampling.\cite{Kmiecik2016} One can learn the atomic-level properties to construct a fast and accurate neural coarse-grained model once the coarse-grained mapping is given. Early attempts to apply DL based methods to coarse-graining focus on water molecules with the roto-translationally invariant features. \cite{zhang2018deepcg, patra2019coarse} \citeauthor{wang2019machine} developed CGNet and learned the coarse-grained model of the mini protein, chignolin, in which the atoms of a residue are mapped to the corresponding C$_\alpha$ atom. The free energy surface learned with CGNet is quantitatively correct and MD simulations performed with CGnet potential predict the same set of metastable states (folded, unfolded, and misfolded). \cite{wang2019machine} Also, the level of coarse-graining, e.g., a single coarse-grained atom to represent a residue versus two coarse-grained atoms, one to represent the backbone and the other to represent the sidechain, is critical to the performance of coarse-grained models. For this purpose, \citeauthor{wanglearning} applied an encoder-decoder based model to explicitly learn the lower-dimensional representation of proteins by minimizing the information loss at different levels of coarse-graining.\cite{wanglearning}

\section{Structure determination}

The most successful application of DL in the field of protein modeling so far has been the prediction of protein structure. Protein structure prediction is formulated as a well-defined problem with clear inputs and outputs: predict the 3D structure (output) given amino acid sequences (input), with the experimental structures as the ground truth (labels). This problem perfectly fits the classical supervised learning approach, and once the problem is defined in these terms, the remaining challenge is to choose a framework to handle the complex relationship between input and output. The CASP experiment for structure prediction is held every two years and served as a platform for DL to compete with state-of-the-art methods and, impressively, outshine them in certain categories. We will first discuss the application of DL to the protein folding problem, and then comment on some problems related to structure determination. Table \ref{tab:structure_prediction} summarizes major DL efforts in structure prediction. 

\subsection{Protein structure prediction}


Before the notable success of DL at CASP12 (2016) and CASP13 (2018), the state-of-the-art methodology employed complex workflows based on a combination of fragment insertion and structure optimization methods such as simulated annealing with a score function or energy potential. Over the last decade, the introduction of co-evolution information in the form of evolutionary coupling analysis (ECA)\cite{Marks2011} improved predictions. ECA relies on the rationale that residue pairs in contact in 3D space tend to evolve or mutate together; otherwise, they would disrupt the structure to destabilize the fold or render a large conformational change. Thus evolutionary couplings from sequencing data suggest distance relationships between residue pairs and aid structure construction from sequence through contact or distance constraints. Since co-evolution information relies on statistical averaging of sequence information from a large number of MSAs,\cite{Ma2015, Skwark2014} this approach is not effective when the protein target has only a few sequence homologs. Neural networks were, at first, introduced to deduce evolutionary couplings between distant homologs, thereby improving ECA-type contact predictions for contact-assisted protein folding.\cite{Marks2011} While the application of neural networks to learn inter-residue protein contacts dates back to the early 2000s,\cite{fariselli2001prediction,horner2007correlated} more recently this approach was adopted by MetaPSICOV (2-layer NN) \cite{jones2014metapsicov}, PConsC2 (2-layer NN) \cite{Skwark2014}, and CoinDCA-NN (5-layer NN) \cite{Ma2015}, which combined neural networks with ECAs. However, there was no significant advantage to neural nets compared to other machine learning methods at that time. \cite{monastyrskyy2014evaluation}

\begin{figure}[h!t]
\centering
\subfigure[Residue Distance Prediction by RaptorX: the overall network architecture of the deep dilated ResNet used in CASP13. Inputs of the first-stage, one-dimensional convolutional layers are a sequence profile, predicted secondary structure and solvent accessibility. The output of the first stage is then converted into a two-dimensional matrix by concatenation and fed into a deep ResNet along with pairwise features (co-evolution information, pairwise contact and distance potential). A discretized inter-residue distance is the output. Additional network layers can be attached to predict torsion angles and secondary structures. Figure from \citeauthor{Xu2019} (2019)\cite{Xu2019}.]{

\begin{minipage}[t]{\linewidth}
\centering
\includegraphics[width=0.9\textwidth]{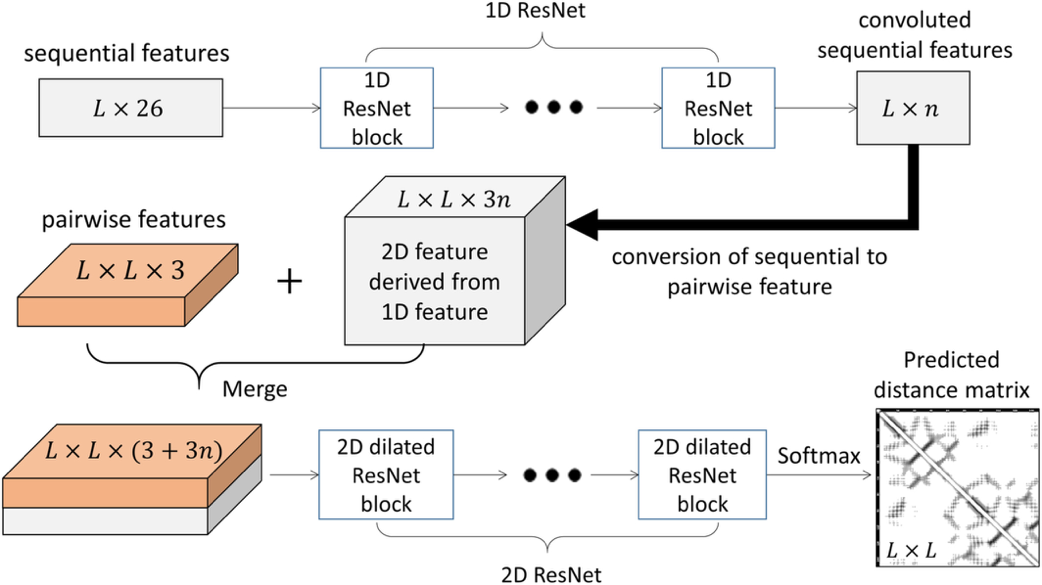}
\end{minipage}%
}
\subfigure[Direct Structure Prediction: Overview of recurrent geometric networks (RGN) approach. The raw amino acid sequence along with a PSSM are fed as input features, one residue at a time, to a bidirectional LSTM net. Three torsion angles for each residue are predicted to directly construct the three-dimensional structure. Figure from AlQuraishi (2019)\cite{AlQuraishi2019}.]{
\begin{minipage}[t]{\linewidth}
\centering
\includegraphics[width=\textwidth]{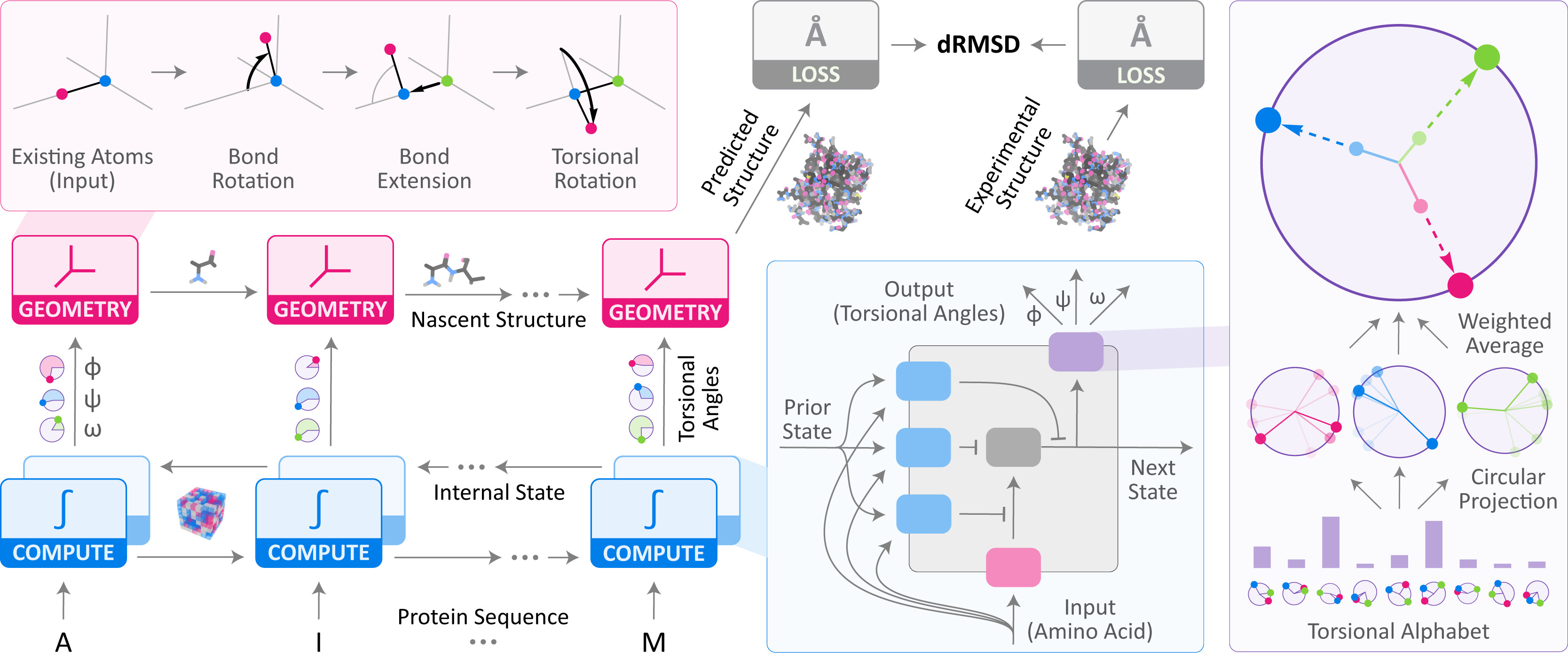}
\end{minipage}%
}
\centering
\caption{Two representative DL approaches to protein structure prediction}
\label{fig:structure_prediction}
\end{figure}

In 2017, \citeauthor{wang2017accurate} \cite{wang2017accurate} proposed RaptorX-Contact, a residual neural network (ResNet) based model, \cite{he2016deep} which, for the first-time employed a \textit{deep} neural network for protein-contact prediction, significantly improving the accuracy on blind, challenging targets with novel folds. RaptorX-Contact ranked first in free modeling (FM) targets at CASP12.\cite{moult2018critical} Its architecture (Figure \ref{fig:structure_prediction} (a)) entails (1) a 1D ResNet that inputs MSAs, predicted secondary structure and solvent accessibility (from DL based prediction tool RaptorX-Property\cite{Wang2016raptoex}) and (2) a 2D ResNet with dilations that inputs the 1D ResNet output and inter-residue co-evolution information from CCMPred\cite{Seemayer2014}. In its original formulation, RaptorX-Contact outputs a binary classification of contacting versus non-contacting residue pairs\cite{wang2017accurate}. Later versions were trained to learn multi-class classification for distance distributions between C$_\beta$ atoms.\cite{xu2019distance} The primary contributors to the accuracy of predictions was the co-evolution information from CCMpred and the depth of the 2D ResNet, suggesting that the deep neural network learned co-evolution information better than previous methods. Later, the method was extended to predict C$_\alpha -$C$_\alpha$, C$_\alpha -$C$_\gamma$, C$_\gamma -$C$_\gamma$, N-O distances and torsion angles (DL based RaptorX-Angle\cite{Gao2018}) and all five distances, torsions, and secondary structure predictions were converted to constraints for folding by CNS\cite{xu2019distance}. At CASP12, however, RaptorX-Contact (original contact based formulation) and DL drew limited attention because the difference between top-ranked predictions from DL-based methods and hybrid DCA-based methods was small.

This situation changed at CASP13 \cite{kryshtafovych2019critical} when one DL-based model, AlphaFold, developed by team A7D, or DeepMind, \cite{senior2020improved, Senior2019, alquraishi2019alphafold} ranked first and significantly improved the accuracy of ``free modeling'' 
(no templates available) targets (Figure \ref{fig:casp13}). The A7D team modified the traditional simulated annealing protocol with DL-based predictions and tested three protocols based on deep neural networks. Two protocols used memory-augmented simulated annealing (with domain segmentation, and fragment assembly) with potentials generated from predicted inter-residue distance distributions and predicted GDT\_TS, \cite{Zemla1999} respectively, whereas the third protocol directly applies gradient descent optimization on a hybrid potential combining predicted distance and Rosetta score. For the distance prediction network, a deep ResNet, similar to that of RaptorX \cite{wang2017accurate}, inputs MSA data  and predicts the probability of distances between $\beta-$carbons. A second network was trained to predict GDT\_TS of the candidate structure with respect to the true or native structure. The simulated annealing process was improved with a Conditional Variational Auto-Encoder (CVAE) \cite{kingma2014semi} model that 
constructs a mapping between the backbone torsions and a latent space conditioned by sequence. With this network, the team generated a database of nine-residue fragments for the memory-augmented simulated annealing system. Gradient-based optimization performed slightly better than the simulated annealing, suggesting that traditional simulated annealing is no longer necessary and state-of-the-art performance can be reached with simply optimizing a network predicted potential. AlphaFold's authors, like the RaptorX-Contact group, emphasized that the accuracy of predictions relied heavily on learned distance distributions and coevolutionary data. 

\citeauthor{yang2019improved} further improved the accuracy of predictions on CASP13 targets using a shallower network than former models (61 versus 220 ResNet blocks in AlphaFold) by additionally training their neural network model (named trRosetta) to learn inter-residue orientations along with $\beta-$carbon distances\cite{yang2019improved} . The geometric features -- C$_{\alpha}$-C$_{\beta}$\ torsions, pseudo-bond angles, and azimuthal rotations -- directly describe the relevant coordinates for the physical interaction of two amino acid side chains.  These additional outputs created significant improvement on a relatively fixed DL-framework, suggesting that there is room for additional improvement. 


An alternative and intuitive approach to structure prediction is directly learning the mapping from sequence to structure with a neural network. \citeauthor{AlQuraishi2019} developed such an end-to-end differentiable protein structure predictor, called Recurrent Geometric Network (RGN), that allows direct prediction of torsion angles to construct the protein backbone (Figure \ref{fig:structure_prediction}b). \cite{AlQuraishi2019} RGN is a bi-directional LSTM that inputs a sequence, PSSM, and positional information and outputs predicted backbone torsions. Overall 3D structure predictions are within 1-2 \AA \  of those made by top-ranked groups at CASP13, and this approach boasts a considerable advantage in prediction time compared to strategies that learn potentials. Moreover, the method does not use  MSA-based information and could potentially be improved with the inclusion of evolutionary information. The RGN strategy is generalizable and well-suited for protein-structure prediction. Several generative methods (see below) also entail end-to-end structure prediction models, like the CVAE framework used by AlphaFold, albeit with more limited success. \cite{Senior2019}

\subsection{Related applications}

\textbf{Protein-Protein Interface} (PPI) prediction identifies residues at the interface of the two proteins forming a complex. Once the interface residues are determined, a local search and scoring protocol can be used to determine the structure of a complex. Similar to protein folding, efforts have focused on learning to classify contact or not. For example, \citeauthor{Townshend2018}  developed a 3D CNN model (SASNet) that voxelizes the three-dimensional environment around the target residue,\cite{Townshend2018} and \citeauthor{fout2017protein} developed a GCN-based model with each interacting partner represented as a graph.\cite{fout2017protein}  Unlike those starting from the unbound structures,\citeauthor{zeng2018complexcontact}  reuse the model trained on single-chain proteins (\textit{i.e.}, RaptorX-Contact) to predict PPI with sequence information alone, which resulted in RaptorX-Complex that outperforms ECA-based methods at contact prediction.\cite{zeng2018complexcontact}  Another interesting approach directly compares the geometry of two protein patches. \citeauthor{gainza2019deciphering}  trained their MaSIF model by minimizing the Euclidean distances between the complementary surface patches on the two proteins while maximizing the distances between non-interacting surface patches.\cite{gainza2019deciphering} This step is followed by a quick nearest neighbor scanning to predict binding partners. The accuracy of MaSIF was comparable to traditional docking methods. However, MaSIF, similar to existing methods, showed low prediction accuracy for targets that involve conformational changes during binding. 

\textbf{Membrane Proteins} (MPs) are partially or fully embedded in a hydrophobic environment composed of a lipid bilayer, and consequently, they exhibit hydrophobic motifs on the surface unlike majority of the proteins that are water soluble. \citeauthor{li2017predicting} used a DL transfer learning framework comprising one-shot learning from non-MPs to MPs.\cite{li2017predicting} They showed that transfer learning works surprisingly well here because the most frequently occurring contact patterns in soluble proteins and membrane proteins are similar. Other efforts include classification of the trans-membrane topology.\cite{tsirigos2015topcons} Since experimental biophysical data is sparse for membrane proteins, \citeauthor{alford2020} compiled a collection of twelve diverse benchmark sets for membrane protein prediction and design for testing and learning of implicit membrane energy models.\cite{alford2020}

\textbf{Loop modeling} is a special case of structure prediction, where most of the 3D protein structure is given, but coordinates of segments of the polypeptide are missing and need to be completed. Loops are irregular and sometimes flexible segments, and thus their structures have been difficult to capture experimentally or computationally.\cite{stein2013improvements, ruffolo2020} So far, DL frameworks based on inter-residue distance prediction (similar to protein structure prediction)\cite{nguyen2017new} and those based on treating the loop residue distances with the remaining residues as an image inpainting problem\cite{li2017protein} have been applied to loop modeling. Recently, \citeauthor{ruffolo2020}  used a RaptorX-like network setup and a trRosetta geometric representation to predict the structure of antibody hypervariable complementarity-determining region (CDR) H3 loops, which is critical for antigen binding.\cite{ruffolo2020}

\section{Protein design}

We divide the current DL approaches to protein design into two broad categories. The first uses knowledge of other sequences (either ``all'' sequenced proteins or a certain class of proteins) to design sequences directly (Table \ref{tab:sequence_generation}). These approaches are well suited to create new proteins with functionality matching existing proteins based on sequence information alone. The second class follows the ``fold-before-function'' scheme and seeks to stabilize specific 3D structures, perhaps but not necessarily with the intent to perform a desired function (Tables \ref{tab:structure_generation} and \ref{tab:inverse_design}). The first approach can be described as function$\rightarrow$sequence (structure agnostic), and the second approach fits the traditional step-wise inverse design: function $\rightarrow$ structure $\rightarrow$ sequence.

\subsection{Direct design of sequence}

Approaches that attempt to design for sequences parallel work in the field of NLP, where an auto-regressive framework is common, most notably, the RNN. In language processing, an RNN model is able to take the beginning of a sentence and predict the next word in that sentence. Likewise, given a starting amino acid residue or a sequence of residues, a protein design model can output a categorical distribution for each of the 20 amino acid residues for the next position in the sequence. The next residue in the sequence is sampled from this categorical distribution, which in turn is used as the input to predict the following one. Following this approach, new sequences, sampled from the distribution of the training data, are generated, with the goal of having properties similar to those in the training set. \citeauthor{Muller2018} \cite{Muller2018} first applied an LSTM RNN framework to learn sequence patterns of anti-microbial peptides (AMPs) \cite{waghu2014camp}, a highly specialized sequence space of cationic, amphipathic helices. The same group then applied this framework to design membranolytic anticancer peptides (ACPs) \cite{grisoni2018designing}. Twelve of the generated peptides were synthesized and six of them killed MCF7 human breast adenocarcinoma cells with at least three-fold selectivity against human erythrocytes. In another application, instead of traditional RNNs, \citeauthor{Riesselman2019} \cite{Riesselman2019} used a residual causal dilated CNN \cite{yu2015multi} in an auto-regressive way and generated a functional single-domain antibody library conditioned on the naive immune repertoires from llamas; though experimental validation was not presented. Such applications could potentially speed-up and simplify the task of generating sequence libraries in the lab.

Another approach to sequence generation is mapping the latent space to the sequence space, and common strategies to train such a mapping include AEs and GANs. As mentioned earlier, AEs are trained to learn a bi-directional mapping between a discrete design space (sequence) and a continuous real-valued space (latent space). Thus, many applications of AEs employ the learnt latent representation to capture the sequence distribution of a specific class of proteins, and subsequently, to predict the effect of variations in sequence (or mutations) on protein function.\cite{ding2019deciphering, Sinai2017, Riesselman2018}  The utility of this learned latent space, however, is more than that. A well trained real-valued latent space can be used to interpolate between two training samples, or even extrapolate beyond the training data to yield novel sequences. One such example is the PepCVAE model.\cite{Das2018} Following a semi-supervised learning approach, \citeauthor{Das2018} trained a VAE model on an unlabeled dataset of $1.7 \times 10^6$ sequences and then refined the model for the AMP subspace using a 15,000-sequence labeled dataset.\cite{Das2018} By concatenating a conditional code indicating if a peptide is antimicrobial, the CVAE framework allows efficient sampling of AMPs selectively from the broader peptide space. More than 82\% of the generated peptides were predicted to exhibit antimicrobial properties according to a state-of-the-art AMP classifier.\cite{Das2018}

Unlike AEs, GANs focus on learning the uni-directional mapping from a continuous real-valued space to the design space. In an early example, \citeauthor{Killoran2017}'s developed a model that combines a standard GAN and activation maximization to design DNA sequences that bind to a specific protein.\cite{Killoran2017}  \citeauthor{repecka2019expanding}  trained ProteinGAN on the bacterial enzyme malate dehydrogenase (MDH) to generate new enzyme sequences that were active and soluble in vitro, some with over 100 mutations, with a 24\% success rate.\cite{repecka2019expanding} Another interesting GAN-based framework is \citeauthor{Gupta2018}'s  FeedBack GAN (FBGAN) that learns to generate complementary DNA sequences for peptides.\cite{Gupta2018} They add a feedback-loop architecture to optimize the synthetic gene sequences for desired properties using an oracle (an external function analyzer). At every epoch, they update the positive training data for the discriminator with high-scoring sequences from the generator so that the score of generated sequences increases gradually. They demonstrated the efficacy of their model by successfully biasing generated sequences towards anti-microbial activity and a desired secondary structure.

\subsection{Design with structure as intermediate}

Within the fold-before-function scheme, one first picks or design a protein fold or topology according to certain desirable properties, then determines the amino acid sequence that could fold into that structure (function $\rightarrow$ structure $\rightarrow$ sequence). The main challenge in generating feasible protein structures might still be choosing a suitable representation of protein structures. \citeauthor{anand2018generative} tested various representations (full atom, torsion-only, etc.) with a deep convolutional GAN (DCGAN) framework that generates sequence-agnostic, fixed-length short protein structural fragments.\cite{anand2018generative}  They found that the distance map of C$_\alpha$ atoms gives the most meaningful protein structures, though the asymmetry of $\psi$ and $\phi$ torsion angles \cite{ramachandran1963stereochemistry} was only recovered with torsion-based representations. Later they extended this work to all atoms in the backbone and combined with a recovery network to avoid the time-consuming structure reconstruction process. \cite{anand2019fully} They showed that some of the designed folds are stable in molecular simulation. Further work in this direction, such as conditioned generation or variable length generation, would enable the design of folds with desired functions.

As for the amino acid sequence design given a protein structure, under the supervised learning setting, most efforts use the native sequences as the ground truth and recovery rate of native-sequences (\textit{i.e.} the percentage of sequence that matches the native one) as a success metric. To compare, \citeauthor{kuhlman2000native} reported sequence-recovery rates of 51\% for core residues and 27\% for all amino acid residues using traditional \emph{de novo} design approaches.\cite{kuhlman2000native} Since the mapping from sequence to structure is not unique (within a neighborhood of each structure), it is not clear that higher sequence recovery rates would be meaningful. A class of efforts, pioneered by the SPIN model \cite{li2014direct}, inputs a five-residue sliding window to predict the amino acid probabilities for the center position to generate sequences compatible with a desired structure. The features in such models include $\phi$ and $\psi$ dihedrals, a sequence profile of a 5-residue fragment derived from similar structures, and a rotamer-based energy profile of the target residue using the DFIRE potential. SPIN  \cite{li2014direct} reached a 30.7\% sequence recovery rate and \citeauthor{wang2018computational}  and \citeauthor{o2018spin2}'s SPIN2  further improved it to 34\%.\cite{wang2018computational,o2018spin2} Another class of efforts inputs the voxelized local environment of an amino acid residue. In \citeauthor{zhang2019prodconn}'s  and \citeauthor{shroff2019structure}'s  models, voxelized local environment was fed into a 3D CNN framework to predict the most stable residue type at the center of a region.\cite{zhang2019prodconn,shroff2019structure} \citeauthor{shroff2019structure}  reported a 70\% recovery rate and the mutation sites were validated experimentally.\cite{shroff2019structure} \citeauthor{anand2020protein}  trained a similar model to design sequences for a given backbone.\cite{anand2020protein} Their protocol involves iteratively sampling from predicted conditional distributions, and it recovered from 33\% to 87\% of native sequence identities. They tested their model by designing sequences for five proteins including a \emph{de novo} TIM-barrel. The designed sequences were 30-40\% identical to native sequences and predicted structures were 2-5 \AA \  RMSD from the native conformation.

Another approach is to generate the full sequence instead of fragments, conditioned by a target structure. \citeauthor{Greener2018} \cite{Greener2018} trained a CVAE model to generate sequences conditioned on protein topology represented in a string\cite{Taylor2002}. The resulting sequence was verified to be stable with molecular simulation. \citeauthor{karimi2019novo}  developed gcWGAN that combined a CGAN and a guidance strategy to bias the generated sequences towards a desired structure. \cite{karimi2019novo} They employed a fast structure prediction algorithm\cite{hou2017deepsf} as an ``oracle'' to assess the output sequence and provide feedback to refine the model. They examined the model for six folds using Rosetta-based structure prediction, and gcWGAN had higher TM-score distributions and more diverse sequence profiles than CVAE.\cite{Greener2018} Another notable experiment is \citeauthor{ingraham2019generative}'s  graph transformer model that inputs a structure, represented as a graph, and outputs the sequence profile.\cite{ingraham2019generative} They treat the sequence design problem similar to a machine translation problem, \textit{i.e.}, a translation from structure to sequence. 
Like the original transformer model\cite{Vaswani2017}, they adopted an encoder-decoder framework with self-attention mechanisms to dynamically learn the relationship between information in two neighbor layers. They measured their results by perplexity, a widely used metric in speech recognition \cite{jelinek1977perplexity}, and the per-residue perplexity (lower is better) for single chains was 9.15, lower than the perplexity for SPIN2 (12.86). 

\section{Outlook and conclusion}

In this review, we have summarized the current state-of-the-art DL techniques applied to the problem of protein structure prediction and design. As in many other areas, DL shows the potential to revolutionize the field of protein modeling. While DL originated from computer vision, NLP and machine learning, its fast development combined with knowledge from operations research, \cite{sutton2018reinforcement} game theory, \cite{goodfellow2014generative} and variational inference \cite{kingma2013auto} amongst other fields, has resulted in many new and powerful frameworks to solve increasingly complex problems. The application of DL for biomolecular structure has just begun, and we expect to see more efforts on methodology development and applications in protein modeling and design.

There are several trends we observed:

\textbf{Experimental validation}: An important gap in current DL work in protein modeling, especially protein design (with few notable exceptions \cite{grisoni2018designing,repecka2019expanding,shroff2019structure}), is the lack of experimental validation. Past blind challenges, \textit{e.g.}, CASP and CAPRI, and design claims have shown that experimental validation in this field is of paramount importance , where computational models are still prone to error. A key next stage for this field is to engage collaborations between machine learning experts and experimental protein engineers to test and validate these emerging approaches.

\textbf{Importance of benchmarking}: In other fields of machine learning, standardized benchmarks have triggered rapid progress. \cite{deng2009imagenet, mayr2016deeptox, brown2019guacamol} CASP is a great example that provides a standardized platform for benchmarking diverse algorithms, including emerging DL-based approaches. A well-defined question and proper evaluation (especially experimental) would lead to more open competition among a broader range of groups and, eventually, the innovation of more diverse and powerful algorithms.

\textbf{Imposing a physics-based prior}: One common topic among the machine learning community is how to utilize existing domain knowledge to reduce the effort during training. Unlike certain classical ML problems such as image classification, in protein modeling, a wide range of biophysical principles restrict the range of plausible solutions. Some examples in related fields include imposing a physics-based model prior \cite{lutter2019deep, greydanus2019hamiltonian}, adding a regularization term with physical meaning, \cite{raissi2019physics} and adopting a specific formula to conserve physical symmetry \cite{zepeda2019deep, han2019universal}. Similarly, in protein modeling, well-established empirical observations can help restrict the solution space, such as the Ramanchandran distribution of backbone torsion angles \cite{ramachandran1963stereochemistry} and the Dunbrack or Richardsons library of side-chain conformations. \cite{shapovalov2011smoothed, hintze2016molprobity} 


\textbf{Closed-loop design}: The performance of DL methodologies relies heavily on the quality of data, but the publicly available datasets may not cover important sample space because of experimental accessibility at the time of experiments. Furthermore, the dataset may contain harmful noise from non-uniform experimental protocols and conditions. A possible solution may be to combine model training with experimental data generation. For instance, one may devise a closed-loop strategy to generate experimental data, on-the-fly, for queries (or model inputs) that are most likely to improve the model, and update the training dataset with the newly generated data.\cite{jensen2019autonomous, coley2019autonomous, coley2019robotic, barrett2019iterative} For such a strategy to be feasible, automated synthesis and characterization is necessary. As high-throughput synthesis and testing of protein (or DNA and RNA) can be carried out in parallel, automation is possible. While such a strategy may seem far-fetched, automated platforms such as those from Ginkgo Bioworks or Transcriptic are already on the market.

\textbf{Reinforcement learning}: Another approach to overcome the limitation of data availability is reinforcement learning (RL). Biologically meaningful data may be generated on-the-fly in simulated environments such as the Foldit game. In the most famous application of RL, AlphaGo Zero, \cite{silver2017mastering} an RL agent (network) was able to learn and master the game by learning from the game-environment alone. There are already some examples of RL in the field of chemistry and electric engineering to optimize the organic molecules or computational chips. \cite{You2018, zhou2019optimization, mirhoseini2020chip} One suitable protein modeling problem for an RL algorithm would be training an AI agent to make a series of ``moves'' to fold a protein, similar to the Foldit game. \cite{cooper2010predicting, koepnick2019novo}

\textbf{Model interpretability}: One should keep in mind that a neural network represents nothing more (and nothing less) than a powerful and flexible regression model. Additionally, due to their highly recursive nature, neural networks tend to be regarded as ``black-boxes'', \textit{i.e.}, too complicated for practitioners to understand the resulting parameters and functions. Although model interpretability in ML is a rapidly developing field, many  popular approaches, such as saliency analysis\cite{zeiler2014visualizing, smilkov2017smoothgrad, sundararajan2017axiomatic} for image classification models, are far from satisfactory.\cite{adebayo2018sanity} Though other approaches\cite{deeplift,SHAP} offer more reliable interpretations, their application to DL model interpretation has been largely missing in protein modeling. As a result, current DL models offer limited understanding of the complex patterns they learn. 


\textbf{The ``sequence $\rightarrow$ structure $\rightarrow$ function'' paradigm}: We know from molecular biophysics that a sequence translates into function through the physical intermediary of a three-dimensional molecular structure. Allosteric proteins\cite{langan2019novo}, for instance, may exhibit different structural conformations under different physiological conditions (\textit{e.g.}, pH) or environmental stimuli (\textit{e.g.} small molecules, inhibitors), reminding us that context is as important as protein sequence. That is, despite Anfinsen's hypothesis,\cite{Anfinsen1973} sequence alone does not always fully determine the structure. Some proteins require chaperones to fold to their native structure, meaning that a sequence could result in non-native conformations when the kinetics of folding to the native structure may be unfavorable in the absence of a chaperone. 
Since many powerful DL algorithms in NLP operate on sequential data, it may seem reasonable to use protein sequences alone for training DL models. In principle, with a suitable framework and training, DL could disentangle the underlying relationships between sequence and structural elements. However, a careful selection of DL frameworks that are structure or mechanism-aware will accelerate learning and improve predictive power. Indeed, many successful DL frameworks applied so far (\textit{e.g.} convolutional neural networks or graph convolutional neural networks) factor in the importance of learning on structural information. 

Finally, with the hope of gaining insight into the fundamental science of biomolecules, there is a desire to link artificial intelligence (AI) approaches to the underlying biochemical and biophysical principles that drive biomolecular function. For more practical purposes, a deeper understanding of underlying principles and hidden patterns that lead to pathology is important in the development of therapeutics. Thus, while efforts strictly limited to sequences are abundant, we believe that models with structural insights will play a more critical role in the future.

\begin{acknowledgement}

This work was supported by the National Institutes of Health through grant R01-GM078221.

\end{acknowledgement}


\begin{sidewaystable}[h] 
\begin{center}
\footnotesize
\caption{Features contained by CUProtein dataset.}
\begin{tabular}{clrlr}
\hline
Feature Name & Description & Dimensions & Type & IO\\
\hline

AA Sequence &
Sequence of amino acid & 
n$\times$1 &
21 chars &
input\\
\hline

PSSM &
Position-specific scoring matrix, a residue-wise score for motifs appearance & 
n$\times$21 &
Real [0, 1] &
input\\
\hline

MSA covariance &
Covariance matrix across homologeousely near sequences & 
n$\times$n &
Real [0, 1] &
input\\
\hline

SS &
A coarse categorized secondary structure (Q3 or Q8) & 
n$\times$1 &
3 or 8 chars &
input\\
\hline

Distance Matrices &
Pairwise distance between residues (C$_{\alpha}$ or C$_{\beta}$) & 
n$\times$ n &
Positive real (\AA) &
output\\
\hline

Torsion Angles &
Variable dihedral angles for each residues ($\phi$, $\psi$) & 
n$\times$ 2 &
Real [-$\pi$, +$\pi$] (radians) &
output\\
\hline

\hline
\end{tabular}
\label{tab:features}
\end{center}
\caption*{n: number of residues in one protein. Data from \citeauthor{drori2019accurate}\cite{drori2019accurate}}
\end{sidewaystable}

\begin{sidewaystable}[h] 
\begin{center}
\caption{A summary of publicly available molecular biology databases.}
\begin{tabular}{clrl}
\hline
Dataset& Description& N& Website \\
\hline

\tabincell{c}{
European Bioinformatics \\
Institute (EMBL-EBI)}& 
\tabincell{l}{
A collections of wide \\
range of datasets}& 
/& 
https://www.ebi.ac.uk\\
\hline

\tabincell{c}{
National Center for \\
Biotechnology Information \\
(NCBI)}& 
\tabincell{l}{
A collections of biomedical \\
and genomic databases}& 
/& 
https://www.ncbi.nlm.nih.gov\\
\hline

\tabincell{c}{
Protein Data Bank \\
(PDB)}& 
\tabincell{l}{
3D structural data of \\
biomolecules, such as\\
proteins and nucleic acids.}& 
$\sim$ 160,000& 
https://www.rcsb.org\\
\hline

\tabincell{c}{
Nucleic Acid Database \\
(NDB)}& 
\tabincell{l}{
Structure of nucleic acids \\
and complex assemblies.}& 
$\sim$ 10,560& 
http://ndbserver.rutgers.edu\\
\hline

\tabincell{c}{
Universal Protein Resource \\
(UniProt)}& 
\tabincell{l}{
Protein sequence and \\ 
function infromations}& 
$\sim 562,000$& 
http://www.uniprot.org/\\
\hline

\tabincell{c}{
Sequence Read Archive \\
(SRA)}& 
\tabincell{l}{
Raw sequence data \\
from ``next-generation'' \\
sequencing technologies}& 
$\sim 3 \times 10^{16} $
NCBI database\\
\hline

\hline
\end{tabular}
\label{tab:data}
\end{center}
\end{sidewaystable}

\begin{sidewaystable}[h] 
\begin{center}
\caption{A summary of structure prediction models.}
\scriptsize
\begin{tabular}{lllllll}
\hline
Model& Architecture & Dataset & N\_train & Performance & Testset & Citation\\
\hline

/ &
MLP(2-layer) &
proteases &
13 &
\tabincell{l}{
3.0 \AA RMSD(1TRM), \\
1.2 \AA RMSD(6PTI)}&
1TRM, 6PTI &
Bohr et al, 1990 \cite{bohr1990novel} \\
\hline

PSICOV &
graphical Lasso &
/ &
/ &
\tabincell{l}{
Precision: Top-L ~0.4, Top-L/2 ~0.53, \\
Top-L/5 ~0.67, Top-L/10 ~0.73}&
150 Pfam &
Jones et al, 2011 \cite{jones2011psicov} \\
\hline

CMAPpro &
2D bi-RNN+MLP &
ASTRAL&
2,352 &
\tabincell{l}{
Precision: Top-L/5 ~0.31, Top-L/10 ~0.4}&
\tabincell{l}{
ASTRAL 1.75 \\
CASP8, 9 }&
Di Lena et al, 2012 \cite{Lena2012} \\
\hline

DNCON &
RBM &
\tabincell{l}{
PDB\\
SVMcon}&
1,230 &
\tabincell{l}{
Precision: Top-L ~0.46, Top-L/2 ~0.55, \\
Top-L/5 ~0.65}&
\tabincell{l}{
SVMCON\_TEST, \\
D329, CASP9}&
Eickholt et al, 2012 \cite{Eickholt2012} \\
\hline

CCMpred &
LM &
\tabincell{l}{
/}&
/ &
\tabincell{l}{
Precision: Top-L ~0.5, Top-L/2 ~0.6, \\
Top-L/5 ~0.75, Top-L/10 ~0.8}&
\tabincell{l}{
150 Pfam}&
Seemayer et al, 2014 \cite{Seemayer2014} \\
\hline

PconsC2 &
Stacked RF &
\tabincell{l}{
PSICOV set}&
150 &
\tabincell{l}{
Positive predictive value(PPV) 0.44}&
\tabincell{l}{
set of 383 \\
CASP10(114)}&
Skwark et al, 2014 \cite{Skwark2014} \\
\hline

MetaPSICOV &
MLP &
\tabincell{l}{
PDB}&
624 &
\tabincell{l}{
Precision: Top-L 0.54, Top-L/2 0.70, \\
Top-L/5 0.83, Top-L/10 0.88}&
\tabincell{l}{
150 Pfam}&
Jones et al, 2014 \cite{jones2014metapsicov} \\
\hline

RaptorX-Contact &
ResNet &
\tabincell{l}{
subset of PDB25}&
6,767 &
\tabincell{l}{
TM-Score: 0.518 \\
(CCMpred: 0.333, MetaPSICOV: 0.377)}&
\tabincell{l}{
Pfam, CASP11, \\
CAMEO, MP}&
Wang et al, 2017 \cite{wang2017accurate} \\
\hline

RaptorX-Distance &
ResNet &
\tabincell{l}{
subset of PDB25}&
6,767 &
\tabincell{l}{
TM-Score: 0.466(CASP12), 0.551(CAMEO), \\
0.474(CASP13)}&
\tabincell{l}{
CASP12+13, \\
CAMEO}&
Xu, 2018 \cite{xu2019distance} \\
\hline

DeepCov &
2D CNN &
\tabincell{l}{
PDB}&
6,729 &
\tabincell{l}{
Precision: Top-L 0.406, Top-L/2 0.523, \\
Top-L/5 0.611, Top-L/10 0.642}&
\tabincell{l}{
CASP12}&
Jones et al, 2018 \cite{Jones2018} \\
\hline

SPOT &
ResNet, Res-bi-LSTM &
\tabincell{l}{
PDB}&
11,200 &
\tabincell{l}{
AUC: 0.958 \\
(RaptorX-contact ranked 2nd: 0.909)}&
\tabincell{l}{
1,250 chains \\
after Jun. 2015}&
Hanson et al, 2018 \cite{Hanson2018} \\
\hline

DeepMetaPSICOV &
ResNet &
\tabincell{l}{
PDB}&
6,729 &
\tabincell{l}{
Precision: Top-L/5 0.6618}&
\tabincell{l}{
CASP13}&
Kandathil et al, 2019 \cite{Kandathil2019} \\
\hline

MULTICOM &
2D CNN &
\tabincell{l}{
CASP 8-11}&
425 &
\tabincell{l}{
TM-Score: 0.69, GDT\_TS: 63.54, \\
SUM Zscore(-2.0): 99.47}&
\tabincell{l}{
CASP13}&
Hou et al, 2019 \cite{Hou2019} \\
\hline

C‐I‐TASSER* &
2D CNN &
\tabincell{l}{
/}&
/ &
\tabincell{l}{
TM-Score: 0.67, GDT\_HA: 0.44, \\
RMSD: 6.19, SUM Zscore($-2.0$): 107.59}&
\tabincell{l}{
CASP13}&
Zheng et al, 2019 \cite{Zheng2019} \\
\hline

AlphaFold &
ResNet &
\tabincell{l}{
PDB}&
31,247 &
\tabincell{l}{
TM-Score: 0.70, GDT\_TS: 61.4, \\
SUM Zscore(-2.0): 120.43}&
\tabincell{l}{
CASP13}&
Senior et al, 2019 \cite{Senior2019, senior2020improved} \\
\hline

MapPred &
ResNet &
\tabincell{l}{
PISCES}&
7,277 &
\tabincell{l}{
Precision: 78.94\% in SPOT, \\
77.06\% in CAMEO, 77.05 in CASP12}&
\tabincell{l}{
SPOT, CAMEO,\\
CASP12}&
Wu et al, 2019 \cite{Wu2019} \\
\hline

trRosetta &
ResNet &
\tabincell{l}{
PDB}&
15,051 &
\tabincell{l}{
TM\_Score: 0.625 (AlphaFold: 0.587)}&
\tabincell{l}{
CASP13, CAMEO}&
Yang et al, 2019 \cite{yang2019improved} \\
\hline

RGN &
bi-LSTM &
\tabincell{l}{
ProteinNet 12 \\
(before 2016)**}&
104,059 &
\tabincell{l}{
10.7 A dRMSD on FM, 6.9 A on TBM}&
\tabincell{l}{
CASP12}&
AlQuraishi, 2019 \cite{AlQuraishi2019} \\
\hline

/ &
bi-GRU, Res LSTM &
\tabincell{l}{
CUProtein}&
75,000 &
\tabincell{l}{
Perseded CASP12 winning team, \\
comparable to AlphaFold in RMSD}&
\tabincell{l}{
CASP12 + 13}&
Drori et al, 2019 \cite{drori2019accurate} \\
\hline

\hline
\end{tabular}
\label{tab:structure_prediction}
\end{center}
\caption*{RBM: restricted Boltzmann machine, LM: pseudo-likelihood maximization, RF: random forest, MLP: multi layer perceptron, MP: membrane protein, FM: free modeling, TBM: template-based modeling \\
* C-I-TASSER and C-QUARK were reported, we only report one here. \\
** RGN was trained on different ProteinNet for each CASP, we report the latest one here.}
\end{sidewaystable}

\begin{sidewaystable}[h] 
\scriptsize
\begin{center}
\caption{Generative models to identify sequence from function (design for function).}
\begin{tabular}{lllllll}
\hline
Model& Architecture& Output & Dataset& N\_train& Performance & Citation\\
\hline

/&
WGAN+AM &
DNA &
\tabincell{l}{
chromosome 1 \\
of human hg38}& 
4.6M &
\tabincell{l}{
$\sim 4$ times stronger than training data \\
in predicted TF binding }& 
Killoran et al, 2017 \cite{Killoran2017} \\
\hline

/ &
VAE &
AA &
\tabincell{l}{
5 protein families}& 
\tabincell{l}{
/}& 
\tabincell{l}{
Natural mutation probability \\
prediction rho=0.58}& 
Sinai et al, 2017 \cite{Sinai2017} \\
\hline

/&
LSTM &
AA &
\tabincell{l}{
ADAM, APD, DADP}& 
1,554 &
\tabincell{l}{
Predicted antimicrobial property 0.79$\pm$0.25\\
(Random: 0.63$\pm$0.26)}& 
Muller et al, 2018 \cite{Muller2018} \\
\hline

PepCVAE &
CVAE &
AA &
\tabincell{l}{
/}& 
\tabincell{l}{
15K labeled, \\
1.7M unlabeled}& 
\tabincell{l}{
Generate predicted AMP with 83\% \\
(random: 28\%, length<30)}& 
Das et al, 2018 \cite{Das2018} \\
\hline

FBGAN &
WGAN &
DNA &
\tabincell{l}{
UniProt (res.<50)}& 
\tabincell{l}{
3,655}& 
\tabincell{l}{
Predicted antimicrobial property \\
over 0.9 after 60 epoches}& 
Gupta et al, 2018 \cite{Gupta2018} \\
\hline

DeepSequence &
VAE &
AA &
\tabincell{l}{
Mutational scan data}& 
\tabincell{l}{
41 scans}& 
\tabincell{l}{
Aimed for mutation effect prediction, \\
outperformed previous models}& 
Riesselman et al, 2018 \cite{Riesselman2018} \\
\hline

DbAS-VAE &
VAE+AS &
DNA &
\tabincell{l}{
simulated data}& 
\tabincell{l}{
/}& 
\tabincell{l}{
Predicted protein expression surpassed \\
FB-GAN/VAE}& 
Brookes et al, 2018 \cite{Brookes2018} \\
\hline

/ &
LSTM &
Musical scores &
\tabincell{l}{
/}& 
\tabincell{l}{
56 betas+\\
38 alphas}& 
\tabincell{l}{
Generated proteins capture the \\
secondary structure feature}& 
Yu et al, 2019 \cite{yu2019self} \\
\hline

BioSeqVAE &
VAE &
AA &
\tabincell{l}{
UniProt}& 
\tabincell{l}{
200,000}& 
\tabincell{l}{
83.7\% reconstruction accuracy, \\
70.6\% EC accuracy}& 
Costello et al, 2019 \cite{costello2019hallucinate} \\
\hline

/ &
WGAN &
AA &
\tabincell{l}{
antibiotic resistance\\
determinants}& 
\tabincell{l}{
6,023}& 
\tabincell{l}{
29\% similar to training sequence(BLASTp)}& 
Chhibbar et al, 2019 \cite{Chhibbar2019} \\
\hline

PEVAE &
VAE &
AA &
\tabincell{l}{
3 protein families}& 
\tabincell{l}{
31,062}& 
\tabincell{l}{
Latent space captures phylogenetic, \\
ancestral relationship, and stability}& 
Ding et al, 2019 \cite{ding2019deciphering} \\
\hline

/ &
ResNet &
AA &
\tabincell{l}{
mutation data+\\
Ilama immune repertoire}& 
\tabincell{l}{
1.2M(nano)}& 
\tabincell{l}{
Predicted mutation effect reached \\
state-of-the-art, built a library of CDR3 seq.}& 
Riesselman et al, 2019 \cite{Riesselman2019} \\
\hline

Vampire &
VAE &
AA &
\tabincell{l}{
immuneACCESS}& 
\tabincell{l}{
/}& 
\tabincell{l}{
Generated sequences predicted to be \\
similar to real CDR3 sequences.}& 
Davidson et al, 2019 \cite{davidsen2019deep} \\
\hline

ProGAN &
CGAN &
AA &
\tabincell{l}{
eSol}& 
\tabincell{l}{
2,833}& 
\tabincell{l}{
Solubility prediction $R^2$ improved \\
from 0.41 to 0.45}& 
Han et al, 2019 \cite{han2019progan} \\
\hline

ProteinGAN &
GAN &
AA &
\tabincell{l}{
MDH from Uniprot}& 
\tabincell{l}{
16,706}& 
\tabincell{l}{
60 sequence were tested in vitro, \\
19 soluble, 13 with catalytic activity}& 
Repecka et al, 2019 \cite{repecka2019expanding} \\
\hline

CbAS-VAE &
VAE+AS &
AA &
\tabincell{l}{
protein fluorescence dataset}& 
\tabincell{l}{
5,000}& 
\tabincell{l}{
Predicted protein fluorescence surpassed \\
FB-VAE/DbAS}& 
Brookes et al, 2019 \cite{Brookes2019} \\
\hline

\hline
\end{tabular}
\label{tab:sequence_generation}
\end{center}
\caption*{AM: activation maximization, CVAE: conditional variational auto-encoder, CGAN: conditional generative adversarial network, AS: adaptive sampling, EC: enzyme commission \\
AA: amino acid sequence, DNA: DNA sequence}
\end{sidewaystable}

\begin{sidewaystable}[h] 
\scriptsize
\begin{center}
\caption{Generative models for protein structure prediction.}
\begin{tabular}{lllllll}
\hline
Model& Architecture& Representation & Dataset& N\_train& Performance & Citation\\
\hline

/ &
DCGAN &
C$_\alpha$-C$_\alpha$ distances &
\tabincell{l}{
PDB (16, 64, 128-residue fragment)}& 
115,850 &
\tabincell{l}{
Meaningful secondary structure, \\
reasonable Ramachandran plot}& 
Anand et al, 2018 \cite{anand2018generative} \\
\hline

RamaNet &
GAN &
Torsion angles &
\tabincell{l}{
Ideal helical structures from PDB}& 
607 &
\tabincell{l}{
Generated torsions are concentrated \\
around helical region }& 
Sabban et al, 2019 \cite{Sabban2019} \\
\hline

/ &
DCGAN &
Backbone distance &
\tabincell{l}{
PDB (64-residue fragment)}& 
800,000 &
\tabincell{l}{
Smooth interpolations. Recover from \\
sequence design and folding. }& 
Anand et al, 2019 \cite{anand2019fully} \\
\hline

\hline
\end{tabular}
\label{tab:structure_generation}
\end{center}
\caption*{
GAN: generative adversarial network \\
DCGAN: deep convolutional generative adversarial network 
}
\end{sidewaystable}

\begin{sidewaystable}[h] 
\scriptsize
\begin{center}
\caption{Generative models to identify sequence from structure (protein design).}
\begin{tabular}{lllllll}
\hline
Model& Architecture& Input & Dataset& N\_train& Performance & Citation\\
\hline

SPIN &
MLP &
Sliding window with 136 features &
\tabincell{l}{
PISCES}& 
1,532 &
\tabincell{l}{
Sequence recovery of 30.7\% \\
on 1,532 proteins (CV)}& 
Li et al, 2014 \cite{li2014direct} \\
\hline

SPIN2 &
MLP &
Sliding window with 190 features &
\tabincell{l}{
PISCES}& 
1,532 &
\tabincell{l}{
Sequence recovery of 34.4\% \\
on 1,532 proteins (CV)}& 
O'Connell et al, 2018 \cite{o2018spin2} \\
\hline

/ &
MLP &
Target residue and its neighbor as pairs &
\tabincell{l}{
PDB}& 
10,173 &
\tabincell{l}{
Sequence recovery of 34\% \\
on 10,173 proteins}& 
Wang et al, 2018 \cite{Wang2018} \\
\hline

/ &
CVAE &
String encoded structure or metal &
\tabincell{l}{
PDB, \\
MetalPDB}& 
3,785 &
\tabincell{l}{
Verified with structure prediction \\
and dynamic simulation}& 
Greener et al, 2018 \cite{Greener2018} \\
\hline

SPROF &
\tabincell{l}{
Bi-LSTM+\\
2D ResNet}& 
112 1-D features+C$_\alpha$ distance map &
\tabincell{l}{
PDB}& 
11,200 &
\tabincell{l}{
Sequence recovery of 39.8\% \\
on proteins}& 
Chen et al, 2019 \cite{chen2019improve} \\
\hline

ProDCoNN &
3D CNN &
Gridded atomic coordinates &
\tabincell{l}{
PDB}& 
17,044 &
\tabincell{l}{
Sequence recovery of 42.2\% \\
on 5,041 proteins}& 
Zhang et al, 2019 \cite{zhang2019prodconn} \\
\hline

/ &
3D CNN &
Gridded atomic coordinates &
\tabincell{l}{
PDB-REDO}& 
19,436 &
\tabincell{l}{
Sequence recovery ~70\%, \\
experimental validation of mutation}& 
Shroff et al, 2019 \cite{shroff2019structure} \\
\hline

ProteinSolver &
Graph NN &
Partial sequence, adjacency matrix &
\tabincell{l}{
UniParc}& 
$72 \times 10^6$ resi. &
\tabincell{l}{
Sequence recovery of 35\%, \\
folding and MD test with 4 proteins}& 
Strokach et al, 2019 \cite{strokach2019designing} \\
\hline

gcWGAN &
CGAN &
Random noise + structure &
\tabincell{l}{
SCOPe}& 
20,125 &
\tabincell{l}{
Diversity and TM-Score of prediction\\
from designed sequence $\ge$ cVAE}& 
Karimi et al, 2019 \cite{Karimi2019a} \\
\hline

/ &
\tabincell{l}{
Graph \\
Transformer}& 
Bakcbone structure in graph &
\tabincell{l}{
CATH based}& 
18,025 &
\tabincell{l}{
Perplexity: 6.56 (rigid), 11.13 \\
(flexible) (Random: 20.00)}& 
Ingraham et al, 2019 \cite{ingraham2019generative} \\
\hline

DenseCPD &
ResNet &
Gridded backbone atomic density &
\tabincell{l}{
PISCES }& 
$2.6 \times 10^6$ resi. &
\tabincell{l}{
Sequence recovery of 54.45\% \\
on 500 proteins}& 
Qi et al, 2020 \cite{qi2020densecpd} \\
\hline

/ &
3D CNN &
Gridded atomic coordinates &
\tabincell{l}{
PDB }& 
21,147 &
\tabincell{l}{
Sequence recovery from 33\% to 87\%, \\
test with folding of TIM-barrel}& 
Anand et al, 2020 \cite{anand2020protein} \\
\hline

\hline
\end{tabular}
\label{tab:inverse_design}
\end{center}
\caption*{
MLP: Multi-Layer Perceptron, Bi-LSTM: Bidirectional Long Short Term Memory
CV: cross validation, resi.: residues
}
\end{sidewaystable}






\begin{singlespace} 
\bibliography{paper}
\end{singlespace}   






\end{document}